\documentclass[longauth]{aa}

\usepackage{graphicx}
\usepackage{txfonts}
\usepackage{natbib}
\usepackage{amsmath}
\usepackage{hyperref}
\usepackage{xcolor,colortbl}
\usepackage{url}
\usepackage{color}
\usepackage{orcidlink}
\usepackage{soul}
\usepackage{todonotes}
\usepackage{lscape}

\usepackage{multirow}
\newcommand{\Msun}{\ensuremath{\,{M}_\odot}}                      
\newcommand{\Rsun}{\ensuremath{\,{R}_\odot}}                      
\newcommand{\Lsun}{\ensuremath{\,{L}_\odot}}                      
\newcommand{\Mjup}{\ensuremath{\,{M}_{\rm Jup}}}                  
\newcommand{\mic}{\ensuremath{{\mu {m}}}\,}
 
\newcommand{\Mearth}{\ensuremath{\,{M}_{\bigoplus}}} 

\def\degr{\hbox{$^\circ$}}

\begin{document}
	
	\title{MIRI-JWST mid-infrared direct imaging of the debris disk of HD106906 }
	\subtitle{Structure and mass of the disk}
	\titlerunning{MIRI imaging of the disk of HD106906}
	\authorrunning{Rouan et al.}
	\author{ 
	Daniel Rouan\orcidlink{0000-0002-2352-1736}\inst{{\ref{lira}}} \thanks{\email{daniel.rouan@obspm.fr} Based on observations collected with JWST through observing proposals 1277, and 1241},
	Anthony Boccaletti\orcidlink{0000-0001-9353-2724}\inst{\ref{lira}}, 
    Clément Perrot\orcidlink{0000-0003-3831-0381}\inst{\ref{lira}},
    Pierre Baudoz\orcidlink{0000-0002-2711-7116}\inst{\ref{lira}}, 
    Mathilde Mâlin\orcidlink{0000-0002-2918-8479}\inst{\ref{jhu},\ref{stsci},\ref{lira}},
    Pierre-Olivier Lagage\inst{\ref{cea}},
    Rens Waters\orcidlink{0000-0002-5462-9387}\inst{\ref{radboud},\ref{hfml},\ref{sron}},
    Manuel G\"udel\orcidlink{0000-0001-9818-0588}\inst{\ref{vienna},\ref{eth}},
    Thomas Henning\orcidlink{0000-0002-1493-300X}\inst{\ref{mpia}},
    Bart Vandenbussche\orcidlink{0000-0002-1368-3109}\inst{\ref{leuven}},
    Olivier Absil\orcidlink{0000-0002-4006-6237}\inst{\ref{star}},
    David Barrado\orcidlink{0000-0002-5971-9242}\inst{\ref{cab}},
    Christophe Cossou\orcidlink{0000-0001-5350-4796}\inst{\ref{parissaclay}}, 
    Leen Decin\orcidlink{0000-0002-5342-8612}\inst{\ref{leuven}},
    Adrian M. Glauser\orcidlink{0000-0001-9250-1547}\inst{\ref{eth}},
    John Pye\orcidlink{0000-0002-0932-4330}\inst{\ref{leicester}},
    Polychronis Patapis\orcidlink{0000-0001-8718-3732}\inst{\ref{eth}},
    Niall Whiteford\orcidlink{0000-0001-8818-1544}\inst{\ref{museum}},
    Eugene Serabyn\inst{\ref{jpl}},
    Elodie Choquet\orcidlink{0000-0002-9173-0740}\inst{\ref{lam}},
    G\"oran Ostlin\orcidlink{0000-0002-3005-1349}\inst{\ref{oskar}},
    Tom P.\ Ray\orcidlink{0000-0002-2110-1068}\inst{\ref{dublin}}
    Gillian Wright\orcidlink{0000-0001-7416-7936}\inst{\ref{ukatc}}    		
			}  	
\institute{
LIRA, Observatoire de Paris, Universit{\'e} PSL, Sorbonne Universit{\'e}, Sorbonne Paris Cit{\'e}, CY Cergy Paris Universit{\'e}, CNRS, 5 place Jules Janssen, 92195 Meudon, France\label{lira} 
\and Space Telescope Science Institute, 3700 San Martin Drive, Baltimore, MD 21218, USA\label{stsci}
\and Department of Physics \& Astronomy, Johns Hopkins University, 3400 N. Charles Street, Baltimore, MD 21218, USA\label{jhu}
\and Universit{\'e} Paris-Saclay, Universit{\'e} Paris Cit{\'e}, CEA, CNRS, AIM, 91191, Gif-sur-Yvette, France\label{cea}
\and  Department of Astrophysics/IMAPP, Radboud University, PO Box 9010, 6500 GL Nijmegen, the Netherlands\label{radboud}
\and  HFML - FELIX. Radboud University PO box 9010, 6500 GL Nijmegen, the Netherlands\label{hfml}
\and  SRON Netherlands Institute for Space Research, Niels Bohrweg 4, 2333 CA Leiden, the Netherlands\label{sron}
\and   Department of Astrophysics, University of Vienna, T\"urkenschanzstrasse 17, 1180 Vienna, Austria\label{vienna}
\and  Max-Planck-Institut f\"ur Astronomie (MPIA), K\"onigstuhl 17, 69117 Heidelberg, Germany \label{mpia}
\and ETH Z\"urich, Institute for Particle Physics and Astrophysics, Wolfgang-Pauli-Strasse 27, 8093 Z\"urich, Switzerland\label{eth}
\and Institute of Astronomy, KU Leuven, Celestijnenlaan 200D, 3001 Leuven, Belgium\label{leuven}
\and STAR Institute, Universit\'e de Li\`ege, All\'ee du Six Ao\^ut 19c, 4000 Li\`ege, Belgium\label{star}
\and Centro de Astrobiología (CAB), CSIC-INTA, ESAC Campus, Camino Bajo del Castillo s/n, 28692 Villanueva de la Cañada, Madrid, Spain \label{cab}
\and LERMA, Observatoire de Paris, Universit\'e PSL, Sorbonne Universit\'e, CNRS, Paris, France\label{lerma}
\and  School of Physics \& Astronomy, 
Space Park Leicester, University of Leicester, 92 Corporation Road, Leicester, LE4 5SP, UK\label{leicester}
\and Department of Astronomy, Stockholm University, AlbaNova University Center, 10691 Stockholm, Sweden\label{stockholm}
\and UK Astronomy Technology Centre, Royal Observatory, Blackford Hill, Edinburgh EH9 3HJ, UK\label{ukatc}
\and Université Paris-Saclay, CEA, IRFU, 91191, Gif-sur-Yvette, France\label{parissaclay}
\and European Space Agency, Space Telescope Science Institute, Baltimore, MD, USA\label{esa}
\and  Kapteyn Institute of Astronomy, University of Groningen, Landleven 12, 9747 AD Groningen, the Netherlands\label{kapteyn}
\and  Institute for Astronomy, University of Edinburgh, Royal Observatory, Blackford Hill, Edinburgh EH9 3HJ\label{roe}
\and Department of Astrophysics, American Museum of Natural History, New York, NY 10024, USA\label{museum}
\and Jet Propulsion Laboratory, California Institute of Technology, 4800 Oak Grove Dr.,Pasadena, CA 91109, USA\label{jpl}
\and Aix Marseille Univ, CNRS, CNES, LAM, Marseille, France\label{lam}
\and Leiden Observatory, Leiden University, P.O. Box 9513, 2300 RA Leiden, the Netherlands\label{leiden}
\and Université Paris-Saclay, UVSQ, CNRS, CEA, Maison de la Simula- tion, 91191, Gif-sur-Yvette, France \label{ups}
\and Cosmic Dawn Center (DAWN), DTU Space, Technical University of Denmark. Building 328, Elektrovej, 2800 Kgs. Lyngby, Denmark\label{dawn}
\and Department of Astronomy, Oskar Klein Centre, Stockholm University, 106 91 Stockholm, Sweden\label{oskar}
\and School of Cosmic Physics, Dublin Institute for Advanced Studies, 31 Fitzwilliam Place, Dublin, D02 XF86, Ireland\label{dublin}
} 
	
	  \abstract
	  { We report MIRI-JWST coronagraphic observations at 11.3 and 15.5 \mic of the debris disk around the young star HD 106906. The wavelength range is sensitive to the thermal
	  emission of the dust heated by the central star.}
	  {The observations were made to characterize the structure of the disk through the thermal emission, to search for clues to the presence of a central void of dust particles, and to derive the mass of the dust and the temperature
	  distribution. Another goal was also to constrain the size distribution of the grains.  }
	  {The data were reduced and calibrated  using the JWST pipeline. The  analysis was based on a forward-modeling of the images using a multiparameter radiative 
	  transfer model coupled to an optical code for coronagraphy processing.}	  
	  {The disk is clearly detected at both wavelengths. The slight asymmetry is 
	  geometrically consistent with the asymmetry observed in the near-IR, but it is inconsistent the brightness distribution. The observed 
	  structure is well reproduced with a model of a disk (or belt) with a critical radius 70 au, a mildly 
	  inward-increasing density (index 2) and a steeper decrease outward (index -6). 
	This indication of a filled disk inside the critical radius is inconsistent with sculpting from an
	  inner massive planet.   
	  The size distribution of the grains that cause the mid-IR emission is  well 
	  constrained by the flux ratio at the two wavelengths : 0.45 -- 10 \mic and 0.65 -- 10  \mic 
	  for silicate and graphite grains, respectively. The minimum size is consistent with predictions 
	  of blowout through radiative pressure. }  	  
	  {We derive a mass of the dust that causes the mid-IR emission of 3.3 -- 5.0\,$10^{-3} \Mearth$.  
	  When the larger grains (up to 1 cm) that cause the  
	  millimeter emission are included, we extrapolate this mass to 0.10 -- 0.16 $\Mearth$. 
	  We point out to that this is fully consistent with ALMA observations of the disk 
	  in terms of dust mass and of its millimeter flux. We estimate the average dust temperature 
	  in the planetesimal belt to be 74 K, but the temperature range within the whole disk is rather wide: from 40 to 130 K. }
	  
	  
	  \maketitle

\section{Introduction}

Debris disks are dusty disks that are most often found around young main-sequence and pre-main-sequence stars. They can help  us to
characterize one important phase in the initial evolution of exoplanetary 
systems. They are produced through collisional cascades, in which 
planetesimals break into millimeter to submicron-sized dust grains (\citealt{wyatt2008, matthews2014, hughes2016}).

Most of debris disks are detected through their far-IR excess, such as the 
archetypical excess around $\beta$\,Pic that was discovered by the IRAS satellite \footnote{Infrared Astronomical Satellite} and then observed in the visible 
(\citealt{aumann1984,smith1984}). Since the advent 
of near-IR instruments with a high angular resolution that use adaptive optics and 
are installed on large telescopes, the direct detection of the 
light that is scattered by grains of the disk became another powerful way to study 
debris disks (\citealt{hughes2018}). The additional benefit of these instruments is that they provide structural information. 
Efficient scattering implies grains of a relatively small size. However, under 
the action of radiation pressure and/or 
stellar winds, dust grains that are smaller than a critical size must be blown out from the 
system on a timescale that is short 
enough to require that the collisional cascade continually produces small  dust 
grains to replenish the disk (\citealt{augereau2006, strubbe2006}). This requires 
that the relative velocity of the planetesimals is sufficient to cause fragmentation upon impact. 
The question of the grain size distribution is then important and is 
directly related to the mechanisms of grain replenishment, blowout, and
fragmentation that occur in the early phase of planet formation. 
Mid-IR resolved imaging is also an efficient  means for studying the grain 
population in debris disks. By probing the thermal emission of the dust, 
it can constrain the size distribution of medium-size grains, their 
distribution in the disk, and their temperature, and it provides a more reliable estimate 
of the total grain mass. 

In the following, we present  observations of the debris disk around the 
binary system HD\,106906 that were obtained using the Mid-Infrared Imager (MIRI) aboard the James Webb Space Telescope (JWST) in its coronagraphic mode. The 
disk is indeed detected in the two filters F1140C (11.3 \mic) and F1550C (15.5 \mic),
and we 
derived several pieces of information on its structure and its luminosity. 
Because of its small angular size, the disk is partly attenuated by the effect 
of the coronagraph. To proceed,  we present an attempt 
to reproduce the observed fluxes and the spatial 
distribution of light using a radiative transfer model. The code takes scattering  and thermal emission as well as a parameterized 
structure of the disk into account. From the comparison of the best-fit model to 
the observations, we derived several important quantities: the grain size 
distribution, the mass of the disk, the temperature map, and the structure.   
Sect. 2 summarizes what is known about the HD\,106906 system. 
Sect. 3 describes details of the observations. 
Sect. 4 provides elements of the data reduction and calibration. 
In Sect. 5 we show the resulting images  at the two 
observation wavelengths. 
Sect. 6 describes the components of the radiative transfer code we used. 
In Sect. 7 we analyze the output of the code for a grid of 96 different sets of 
parameters and compare them to observations to determine the best-fit model of the disk and
the grain population. 
In Sect. 8 we discuss the results for the structure of the disk, the mass, and the temperature of the disk. We also discuss this in terms of consistency with other observations and theoretical works.         
   
	\section{HD106906}
\label{sec:HD106906}	
HD\,106906 (HIP\,59960) is a pre-main-sequence F5V-type binary  stellar 
system (\citealt{Absil2021, derosa2019}; separation: 0.14\,au , P.A.: 95.2\degr)  \, 
featuring a debris disk and an 11\,$\Mjup$ planet companion
\citep{bailey2014} at an 
unusually large projected separation of $\approx$ 800 au. \cite{lagrange2016}
estimated a physical distance between 2000 and 3000 au based on the assumption that the 
planet orbit is coplanar with the disk. The system is a member of 
the lower Centaurus Crux (LCC) association, which has a mean age of 17 $\pm$ 
5 Myr, which agrees with  an isochronal age and mass of 13$\pm$2 
Myr and 1.35 \Msun \, for each component of the binary star (\citealt{pecaut2012, rodet2017}). 
The debris disk was first detected by the Spitzer satellite based on its 
infrared excess \citep{chen2005} and was later observed at near-IR wavelength 
by \cite{kalas2015} with the Gemini Planet Imager (GPI), by \cite{lagrange2016} (quoted as L16 in the 
following) using the instrument Spectro Polarimetric High contrast Exoplanet REsearch (SPHERE) on the Very Large Telescope (VLT), and by \cite{crotts2021} in polarimetry in the 
J, H, and K$_s$ bands. The disk is elongated in the SE-NW direction and is seen nearly 
edge-on. It extends beyond 100\,au, and its brightness distribution peaks at 
about 65--75\,au. 
Using a geometrical model to account for scattering, L16 derived an inclination 
of $85.3 \pm 0.1^{\circ}$ and a disk position angle (PA) of $104.4 \pm 0.3^{\circ}$. 
More recently, 
\cite{kral2020} and \cite{fehr2022} detected and resolved the 
thermal emission of the disk at the millimeter wavelength using the Atacama Large Millimeter/submillimeter Array (ALMA), but did not detect CO.
	\begin{figure}
	   	\centering
	   	\resizebox{\hsize}{!}{\includegraphics{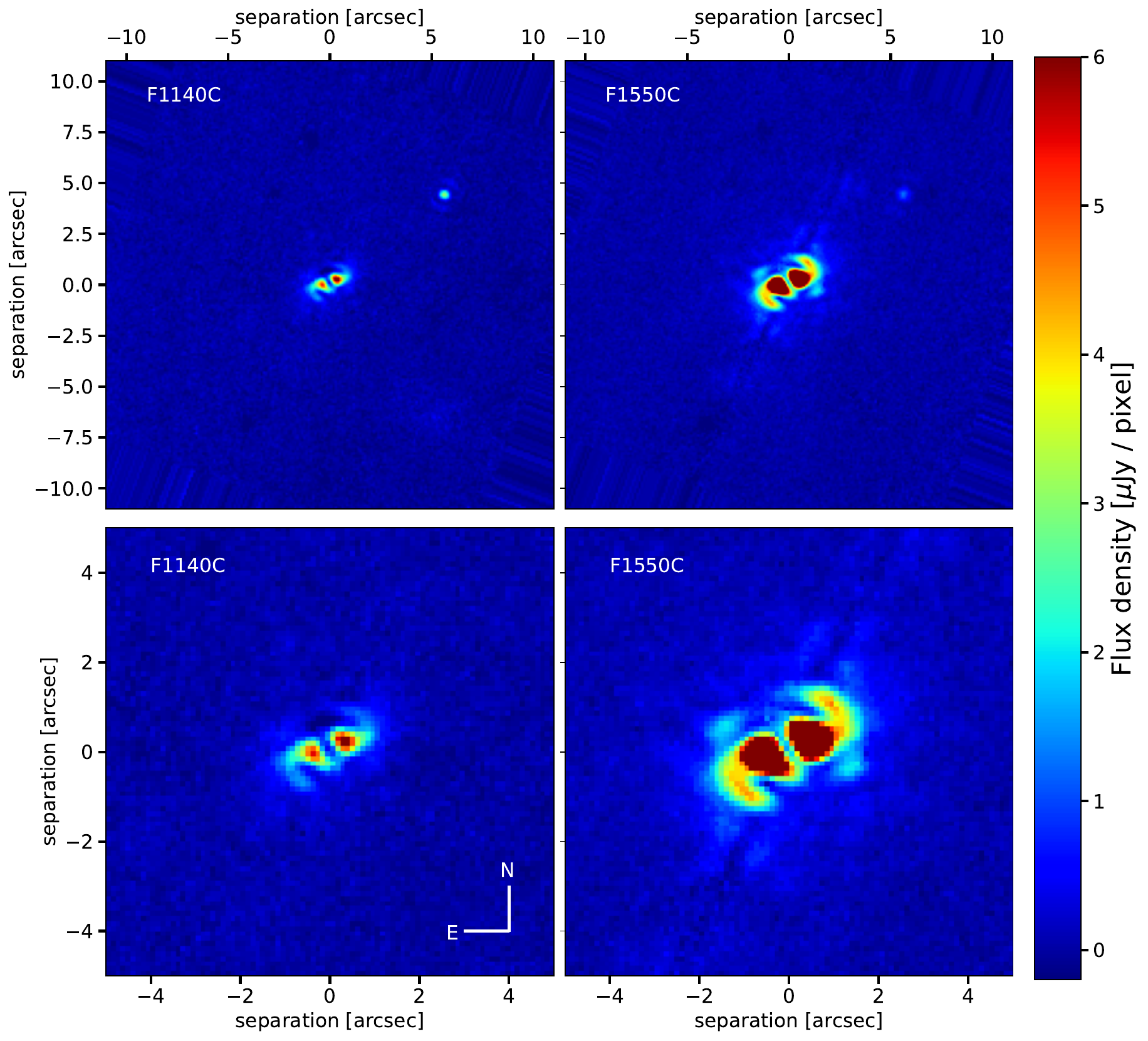}}
	   	\caption{Top: MIRI F1140C (left) and  F1550C (right) 
	   	full-field coronagraphic images after data reduction (see text) and 
	   	reference subtraction. The planet HD\,106906\,b can be seen at 7.3 
	   	arcsec NW. Bottom: Same after zooming and binning by a factor 2. 
	   	The orientation is standard (north is up, and east is to the left). The field of view is 22 $\times$ 22 arcsec$^2$ at the top and 11 $\times$ 11 
	   	arcsec$^2$ at the bottom. }
	   	\label{fig:F1}
   	\end{figure}

\section{ Observations} \label{sec:obs}
\label{sec:observations}

HD\,106906 is included in MIRIco, which is an EU and US coordinated observing effort 
that makes use of MIRI (\citealt{wright2023}) in its coronagraphic mode throughout programs 1277 and 1241.
The target was observed in one run on May 16, 2023, under  Guaranteed 
Time Observations  (GTO) program 1277, using the MIRI Four Quadrant Phase Mask 
(\citealt{rouan2000}) coronagraphs (4QPM) with filters F1140C and F1550C. 
\begin{table*}

\caption{Log of observations.}

\begin{tabular}{llllllllll}
\hline
\hline
Date and time      &  Obs ID &  Filter & Type & Ngroup & Nint & Ndither & Texp \\
UT &&&&&&& [s] \\
\hline

May 16, 2023 18:11:50 & 19 & F1140C &  Target & 500 & 3 & 1 & 360\\
May 16, 2023 19:31:23 & 20 & F1550C &  Target & 500 & 15 & 1 & 1801\\
May 16, 2023 20:26:25 & 21 & F1550C &  Background & 500 & 8 & 2 & 1920\\
May 16, 2023 21:14:54 & 22 & F1140C &  Background & 500 & 2 & 2 & 480\\ \hline
\end{tabular}
\label{tab:Table1}

\tablefoot{Observation parameters of HD\,106906 system as part of the MIRI GTO program 1277. The date and time represent the starting time of the observation on the target, the ID of each observation, the filter, and 
the type of observation. The last parameters represent the observational parameters: The number of groups, the number of integrations, the number of dither positions, and the total exposure time.}
\end{table*}

The goal of the program is to 
characterize the planet and the disk. This paper is essentially devoted to 
the study of the disk, and a forthcoming paper will focus on the planet photometry. The observation log is provided in Table \ref{tab:Table1}. For 
each coronagraphic filter, we observed the target and its 
associated background back to back (in two dithers). The background images were used to remove 
the glowstick effect, as explained by \cite{boccaletti2022}, and they were 
obtained near the target (typically, a few dozen arcseconds away). 
Unlike for the observations of HR\,8799 and 
HD\,95086 (\citealt{boccaletti2024, malin2024}), we did not observe a dedicated 
reference star. 
The reference star we used here to subtract the residues is HD\,218261. This star was also used by \cite{boccaletti2024} for HR\,8799 and was observed in 
November 2022, so there is a gap of almost 6 months. The stability of JWST is 
good enough for the subtraction to be fully efficient, however. The reference has a 
comparable magnitude to HD\,106906 in the MIRI coronagraphic filters. 
Taking into account the stellar residuals and the background noise from the 
diffraction model of \cite{boccaletti2015}, we determined the exposure times to achieve signal-to-noise ratios (S/N) higher than $\approx$ 10 
on the planet. 
The total exposure times on target were 360\,s and 1800\,s  for the  F1140C and 
F1550C filter, respectively.
	
\section{Data reduction and calibration}
\label{sec:reduc}
We retrieved the processed data from the Mikulski Archive for Space 
Telescopes (MAST\footnote{MAST: \url{mast.stsci.edu}}), and we also reprocessed the raw 
data  on our side with v12 of the JWST pipeline\footnote{\url{jwst-pipeline.readthedocs.io}} 
\citep{bushouse2025}
for comparison. The main steps of the process were the same as in \cite{boccaletti2024, malin2024}. 
In brief, stage 1 of the JWST pipeline 
applies ramps-to-slopes processing and delivers count-rate images 
corrected for various detector artifacts 
(linearity, saturation, outliers, etc.). 
Stage 2 is reduced to only applying the 
background subtraction without any flat correction  to avoid increasing noise and the 
glowstick effect. 
We verified that the impact on the photometry is lower than 2\%, which is much smaller 
than the other sources of noise \citep{boccaletti2024}.
In the end, we  collapsed all integrations (Nint) into a single frame to obtain a number of frames that was the 
number of dither positions (Ndither). 
A mean combination of the background dithers was subtracted from each target 
observation, and the remaining bad pixels were rejected with a $\sigma$ 
clipping. 

Each coronagraphic image was registered at the 4QPM centers. This was 
determined during commissioning (using a cross-correlation with a large 
database of simulated data).

Raw coronagraphic images are essentially dominated by the 
diffraction pattern proper to the hexagonal pupil of the JWST. The optimized Lyot 
stop cannot completely suppress this (\citealt{boccaletti2022}).  
We performed reference differential imaging (RDI) with a single reference star and
considered several approaches, including PCA and linear combination as in 
\citet{boccaletti2024} and \citet{malin2024}. As a result, we selected the simplest 
RDI method, which uses a single image of the reference star obtained at one 
out of the nine positions of the small grid dither. This position and the intensity scaling were 
chosen to minimize the diffraction residuals while maximizing the disk intensity.

\section{Resulting images} 
\label{sec:results}
Figure \ref{fig:F1} displays the 
reduced images resulting from the process  described above at the two wavelengths 11.4 and 15.5 $\mu m$  and the same 
after zooming on the central part and binning by a factor of two. 
The disk clearly appears as two elongated patches in the processed images that 
bracket a void that is caused by the coronagraph extinction of the central 
part of the disk. The direction joining the two centers of gravity of the 
patches has a PA of 112\degr\,  , which is
significantly larger than the PA measured in near-IR (104.4\degr\, for 
L16, 103.7\degr\, for \cite{kalas2015}, 103\degr\, for \cite{fehr2022}). 
We  note that a first reason for a difference between near-IR and mid-IR 
orientation of the disk is that it  appears rather as an arc in the near-IR because 
only the scattering edge is seen.
Second, we point out that this difference is not expected to be related to the 
difference between mechanisms of emission that are at play in each wavelength domain 
(scattering versus thermal emission). 
This appears to be obvious when the final images produced by the model are analyzed. 
(see section \ref{sec:model}). They feature the same pattern as in the observed 
image, where the line joining the two lobes has the same orientation, while the model 
produces an initial image (before processing by the coronagraph) that is fully symmetric 
with respect to the imposed direction of PA 104.4\degr. 
We argue that this is the result of  some diffraction effects induced by the four-quadrant 
coronagraph mask, and we note that the direction with respect to the north of the 
frontier between the quadrants is 4.83\degr\, and that this must have some effect on 
the polar angle of the observed structures in the resulting image.   
 
In addition to the patches, two fainter hook-like structures are observed. They 
start from one side of each patch and describe a centro-symmetric 
pattern analogous to spiral arms. We discuss this peculiar feature further below and 
show that it provides a good indicator for constraining the radial structure 
of the disk. We note that the two patches are not rigorously symmetric, and we discuss
this point in Sect. \ref{sec:discuss}.

\section{The disk model}
\label{sec:model}
\begin{figure}
	   	\centering
	   	\resizebox{\hsize}{!}{\includegraphics{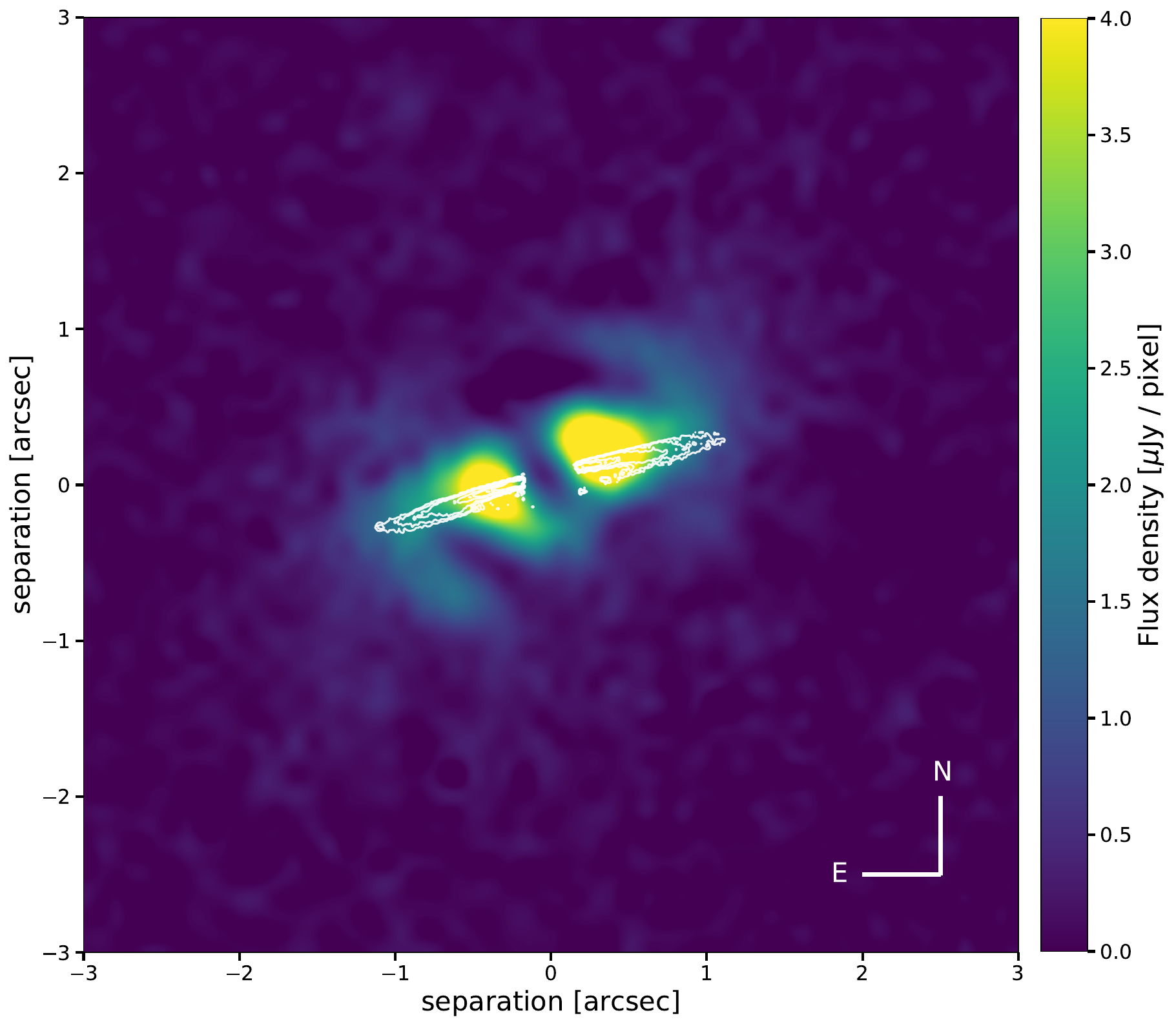}}
	   	\caption{MIRI 11.4 $\mu m$ coronagraphic image. The contours of 
	   	the SPHERE H1-H2 image of Lagrange et al. (2015) are superimposed.}
	   	\label{fig:F2}
   	\end{figure}

The appearance of the disk in the mid-IR is substantially different from  what 
is observed in the near-IR wavelength in scattered light, as shown in Fig. 
\ref{fig:F2}, where the contour of the SPHERE image at H 
(\citealt{lagrange2016}) is superimposed on the MIRI 11.4 $\mu m$ coronagraphic 
image. 
The differences arise from three factors: the lower angular resolution, the 
attenuation of the 4QPM in its transition directions, and the thermal emission 
compactness, which concentrated the observed light in the parts closest to the 
star.
It is then much more difficult than at near-IR to interpret  MIRI images 
straightforwardly in terms of structure, extension, and inclination of the 
disk.    
	  
In order to constrain the structure of the disk and to account for the measured 
fluxes in the two filters, we proceeded by forward-modeling a  physical 
model of the disk through the diffraction model of the MIRI coronagraph. 
We started with the code DDiT described by \cite{olofsson2020} and modified it 
to include several features that were not present in the open-source Python 
version that is available on GitHub\footnote{Available at  https:// github.com/joolof/ 
DDiT} .
Essentially, all the parts of the code concerning the propagation of light 
from the star to the disk and from the disk to the observer are preserved, 
and this also holds for the dust density calculation using the parameterized distribution 
proposed by \cite{augereau1999}. 
For our purpose, we added
\begin{itemize}
\item  the use of wavelength-dependent optical properties of grains, that is, 
essentially, the tables from \cite{laor1993} for Q$_{abs}$, Q$_{sca}$,  and 
$g_{sca}$;
\item the distribution of the grain sizes. We only used the 
classical  a$^{-3.5}$ power law. This law is
expected from an equilibrium collisional cascade 
(\citealt{dohnanyi1968}) and was also proposed for the interstellar medium by   
\cite{mathis1977} (the so-called  
MRN law;  the code can  manage any power law);
\item  a true computation of the radiative equilibrium  temperature of  
grains and of the corresponding thermal emission. For that purpose, we solved for each grain with a radius 
$a$  the radiative equilibrium equation
\(\int Q_{abs}^{a}(\lambda) \, J(\lambda) \,d\lambda = \int Q_{abs}^{a}(\lambda) \, 
4\pi \,B(\lambda)\,d\lambda\), where $J(\lambda)$ is the mean radiation intensity at wavelength $\lambda$, and $B(\lambda)$ is the Planck function;
\item the radius and temperature of the star in order to 
produce a  realistic stellar spectrum; 
\item an actual value of the returned fluxes (in $\mu$Jy) for 
scattered light and  thermal radiation instead of proportional quantities. 
\end{itemize}      

The code (which we refer to as DDiT+ in the following) produces maps of the flux as 
well as maps of the averaged temperature on the line of sight. 

The parameters that were used or adjusted in the simulations were 
 the size of the critical radius of the disk, $\rm R_{c}$;
the distance in parsec; 
the wavelength;
the minimum and maximum grain radii, $a_{min}$ and  $a_{max}$.
As regards the nature of grains (silicate or graphite), we favored silicate grains for two reasons: 
{\it i)} as noted by \cite{Hugues2018} ,
Spitzer spectroscopy of a large sample of debris disks revealed that their dust composition is dominated by standard silicates, and {\it ii)} the Spitzer spectrum of HD106906 (https://cassis.sirtf.com/atlas/) reveals two features in emission at 23.7 and 34.0 \mic that likely are characteristic of forsterite crystalline silicate \citep{Vandenbussche2004} (see Fig. \ref{fig:A5}, where the features are tentatively identified on the spectrum). Note that we also 
considered graphite grains in order to assess the sensitivity of the model results
to the grain nature.  We also adjusted
the $\alpha_{in}$ and $\alpha_{out}$ exponents of the surface density as described by \cite{augereau1999};
the inclination;
the position angle;
the mass of the disk;
the radius and effective temperature of the star (which we considered to be a 
unique star).

The eccentricity was set to zero. We mainly varied the parameter  $\alpha_{in}$ in the disk structure, which sets 
the contribution of the innermost part of the disk. Roughly speaking, 
when $ \alpha_{in}$ = 10, the slope of the density variation is extremely 
steep, and the disk is essentially considered to be a ring at the 
critical radius, while with $\alpha_{in}$ = 1, the density increases continuously from the center to the critical radius. This results in a filled 
disk. 
When scattered light at near-IR was modeled, the slope of the
outer radius was generally considered to be rather steep \citep{kalas2015, lagrange2016}. We 
therefore varied  $\alpha_{out}$ between -4 and -10 
in the simulations.  Only \cite{crotts2021} proposed a fainter slope of 
-2.26, which was based on radiative transfer modeling.
The value -4 was generally 
considered when the external part was dominated by blowout.

We divided the range of grain sizes into bins (typically, 10 to 20), and we computed for each 
bin the fraction of the total mass to which this bin corresponded. We 
ran the model using the proper grain optical properties. We then computed the sum 
of all intensities weighted by the corresponding fraction of mass to obtain 
the final intensity map. 
This was done for the scattered light and for the thermal emission by the 
disk.     
	
The question of the dust size distribution law in HD\,106906 was discussed before. For instance, \cite{crotts2021} considered a power law 
with an index q =  -3.19.  
As noted by \cite{farhat2023}, this distribution differs only slightly from that of a collisional cascade, which is characterized by   $q
\approx$ -3.5 (\citealt{dohnanyi1968}). 
We considered medium-size grains, which are less involved in near-IR 
scattering, and we therefore chose to retain the collisional cascade index.  

The mass of the disk plays no role in the calculation, except for 
the value of the flux, which is directly proportional to the mass. For this reason, 
all our computations were made with a mass of 1 $\Mearth$. By comparing 
the resulting total flux to the  measured flux, we derived the mass of the 
disk for the considered grain population. 
This is the minimum mass because larger grains are probably present, but 
do not contribute to the flux at mid-IR wavelengths, as we discuss below. 

After we produced the final intensity map, we used it as input to a code that 
simulated the effect of the optical system 
of the JWST + MIRI coronagraphic channel as realistically as possible. It took the actual 
shape and aberrations of the primary mirror, the shadow by the secondary struts, 
the effect of the 4QPM, and the optimized Lyot stop into account. Because the disk is somewhat fainter than the star, this simulation of synthetic disk images did 
not include any potential broadening through electronic effects, which is referred to as the 
brighter-fatter effect \citep{Argyriou2023}. This effect appears to be negligible.
The output image was then compared to the  observed image, and we evaluated 
the quality of the fit by computing the reduced
$\chi_{\nu}^2$ between the data and the model. 
The complete modeling (radiative transfer plus optical system) allowed us to estimate
the attenuation by the coronagraph channel.
The attenuation depends on the object and is radically different for a point source
\citep[as estimated in][]{boccaletti2024} 
and a central extended object. For 
instance, we measured a total attenuation for the range of models that varied from 
$\sim$1.8 to 3.4 at F1140C. 

Several parameters were fixed once for all, such as the distance (103 pc), 
because it is strongly constrained by GAIA-DR3 \citep{gaia2023}; the position angle
and the inclination because 
they are well determined by near-IR studies (\citealt{kalas2015, lagrange2016, 
crotts2021}); the radius; and the effective temperature of the star, whose 
spectral type is well determined.  We fixed the effective temperature at 6900 K and the
radius at 1.7 \Rsun, so that the total bolometric luminosity of 5.9 \Lsun \ was similar to the theoretical luminosity of two identical F5V stars  \citep{derosa2019}.   
The  analyses of the structure  of the disk  was based   on 
models using silicate grains only.
 
To discuss the disk mass, we also
considered graphite grains (see Sect.\ref{sec:diskmass}). 

\section{Comparing model and observations}
\label{sec:comparison}
\subsection{Building a grid of models}
\label{subsec:grid}
Using rather conservative parameters taken from near-IR studies and a broad 
grain size range (0.1 - 10 $\mic$), we first compared the scattered and 
thermal intensities to conclude that the former was always several orders of 
magnitude lower than the latter at the wavelength of interest for MIRI coronagraphy. 
Therefore, we only consider the thermal component in the following.
Several parameters can be adjusted, and we therefore adopted the 
following procedure: To determine the best morphological fit between 
the simulated coronagraphic images and the observed images, we first considered a grid 
of 96 (2 $\times$ 3 $\times$ 4 $\times$ 4) models that each corresponded to a 
combination  of four parameters: $\lambda$, $R_{c}$, $\alpha_{in}$, and $
\alpha_{out}$. We then computed the disk image as an input to the 
coronagraphic numerical process for each (see Fig. \ref{fig:coronomodel}) and compared 
the output to the actual observed image with a $\chi^2$ metrics. 
When a good agreement was found, we varied different grain size ranges 
in a second step, until  $a_{min}$ and  $a_{max}$ matched 
the observed ratio $F_{15.5} / F_{11.4}$ well, that is, until they reached a value in the middle of 
the uncertainty range, as shown in Fig.\ref{fig:F7}.
Table \ref{tab:Table2} gives the complete set of values we used for each of these 
parameters to build the grid of models.

\begin{table}

\caption{List of parameter values we used to build the grid of models. }

\begin{tabular}{llllll}
\hline \hline 
Parameter      & \multicolumn{1}{c}{Unit} & \multicolumn{4}{c}{Values}  \\                                                        
\hline
$R_c$          & au   & \multicolumn{1}{l}{65}   & \multicolumn{1}{l}{70}   & \multicolumn{1}{l}{75}\\ 
$\lambda$      & $\mu$m  & \multicolumn{1}{l}{11.4} & \multicolumn{1}{l}{15.5} & \multicolumn{1}{l}{}\\   
$\alpha_{in}$  & --  & \multicolumn{1}{l}{1}    & \multicolumn{1}{l}{2}    & \multicolumn{1}{l}{4}  & 6 \\  
$\alpha_{out}$ & --  & \multicolumn{1}{l}{-4}    & \multicolumn{1}{l}{-6}    & \multicolumn{1}{l}{-8}  & -10 \\ 
inclination & deg & 85 \\
eccentricity & -- & 0 \\
position angle & deg & 104 \\
distance & pc & 103 \\
size distribution power & -- & -3.5 \\
\hline 
\end{tabular}
\label{tab:Table2}

\end{table}
 
\begin{figure*}
	   	\centering

\includegraphics[width=16cm]{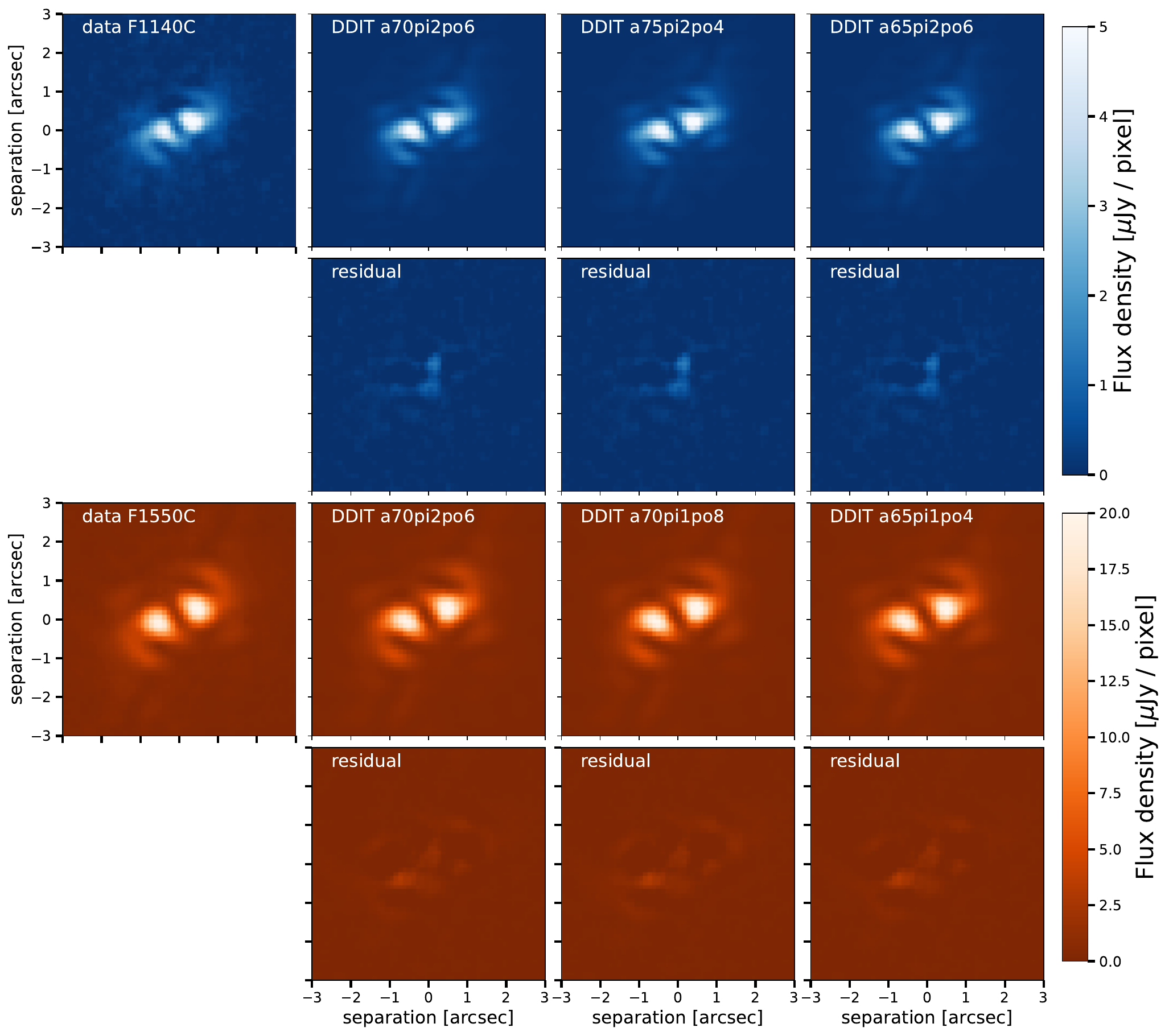}
	   	\caption{Comparison of simulated coronagraphic images and the observed image for a 
	   	few cases for which the set of parameters $\alpha_{in}$, $\alpha_{out}$ , and $R_{c}$ 
	   	led to a rather good solution, as indicated by the low residue level (second 
	   	row of each set). The two upper rows correspond to 
     the F1140C filter, 
     and the two bottom rows show the F1550C filter. 
     The label at the top of each image lists the values of 
	   	the parameters in short. 
For example, a65pi1po4 stands for 
     $R_c$ = 65 au, $\alpha_{in}$ = 1, and $
	   	\alpha_{out}$ = -4. 	
	   	 }
	   	\label{fig:F3}
   	\end{figure*}

A peculiar feature in the observed images played an important role 
in refining the fit: the hook-like structures  
previously mentioned that depart from each lobe. 
They correspond to some diffraction effects, and only certain sets of parameters 
produced their shape and location correctly. We were able to reproduce 
them only for a rather narrow range of $R_{c}$ and of $\alpha_{in}$. As an illustration, 
we display in Fig. \ref{fig:F3} the observed images,  different results (among the best) 
of the simulation for different sets of $\lambda$, $R_{c}$, $\alpha_{in}$, and 
$\alpha_{out}$, and the residue maps. 
The simulation clearly provide fairly similar results to the observed images. 

\begin{table}[]
\caption{Parameters for the set of best models at 15.50\,$\mic$ and 11.40\,$\mic$.}
\begin{tabular}{llllll}
\hline\hline

$\lambda$ & $R_c$ (au), & $\chi_{\nu}^2$ & ratio &  Flux 1  &  Flux 2 \\ 
$\mu$m & $\alpha_{in}$, -$\alpha_{out}$ & && ($\mu$Jy)&  ($\mu$Jy) \\
\hline
\multirow{13}{*}{15.5 }  
& 70, 1, 8  & 3.69  & 3.12  & 7702.6    & 7126.5 \\ 
& 75, 1, 10 & 3.73  & 3.16  & 7667.2    & 7210.7 \\
& 65, 1, 8  & 3.80  & 3.23  & 7752.7    & 7367.2 \\  
& 65, 1, 6  & 3.83  & 3.08  & 7704.2    & 7034.5 \\ 
& 70, 1, 10 & 3.94  & 3.26  & 7773.9    & 7432.2 \\ 
& 75, 1, 8  & 4.39  & 3.03  & 7628.2    & 6909.8 \\ 
& 70, 1, 6  & 4.50  & 3.01  & 7630.7    & 6870.3 \\ 
& \textbf{70, 2, 6}  & \textbf{4.50}  & \textbf{3.01}  & \textbf{7630.7}    & \textbf{6870.3} \\ 
& 65, 1, 4  & 4.52  & 2.99  & 7616.6    & 6830.7 \\ 
& 65, 2, 8  & 4.59  & 2.79  & 6751.4    & 6357.1 \\ 
& 70, 2, 10 & 4.64  & 2.79  & 6720.4    & 6362.0 \\ 
& 65, 2, 10 & 4.76  & 2.92  & 6830.9    & 6664.0 \\ 
& 65, 1, 10 & 4.87  & 3.39  & 7792.9    & 7734.4 \\ 

\hline

\multirow{9}{*}{11.4 }
& 75, 2, 4    & 2.83  & 2.48  & 1204.5    & 1196.7 \\    
& 70, 2, 4    & 2.85  & 2.52  & 1209.7    & 1220.0 \\
& 65, 2, 4    & 2.88  & 2.58  & 1242.6    & 1246.1 \\
& 75, 2, 6    & 2.89  & 2.48  & 1203.1    & 1199.9 \\
& \textbf{70, 2, 6}    & \textbf{2.94}  & \textbf{2.54} &\textbf{1216.0}    & \textbf{1225.4} \\
& 75, 2, 8    & 3.11  & 2.52  & 1206.0    & 1216.4 \\
& 65, 2, 6    & 3.23  & 2.61  & 1242.3    & 1259.7 \\
& 70, 2, 8    & 3.35  & 2.59  & 1236.2    & 1251.1  \\  
\hline
\end{tabular}
\tablefoot{
We display for each wavelength the combination of the model parameters, the $\chi_{\nu}^2$, the coronagraphic attenuation, the flux estimated with method 1, and the flux estimated with method 2.

The unique set of parameters common to the two wavelengths is indicated in bold. }
\label{tab:Table3}
\end{table}

\begin{figure}
	   	\centering
	   	\resizebox{\hsize}{!}{\includegraphics{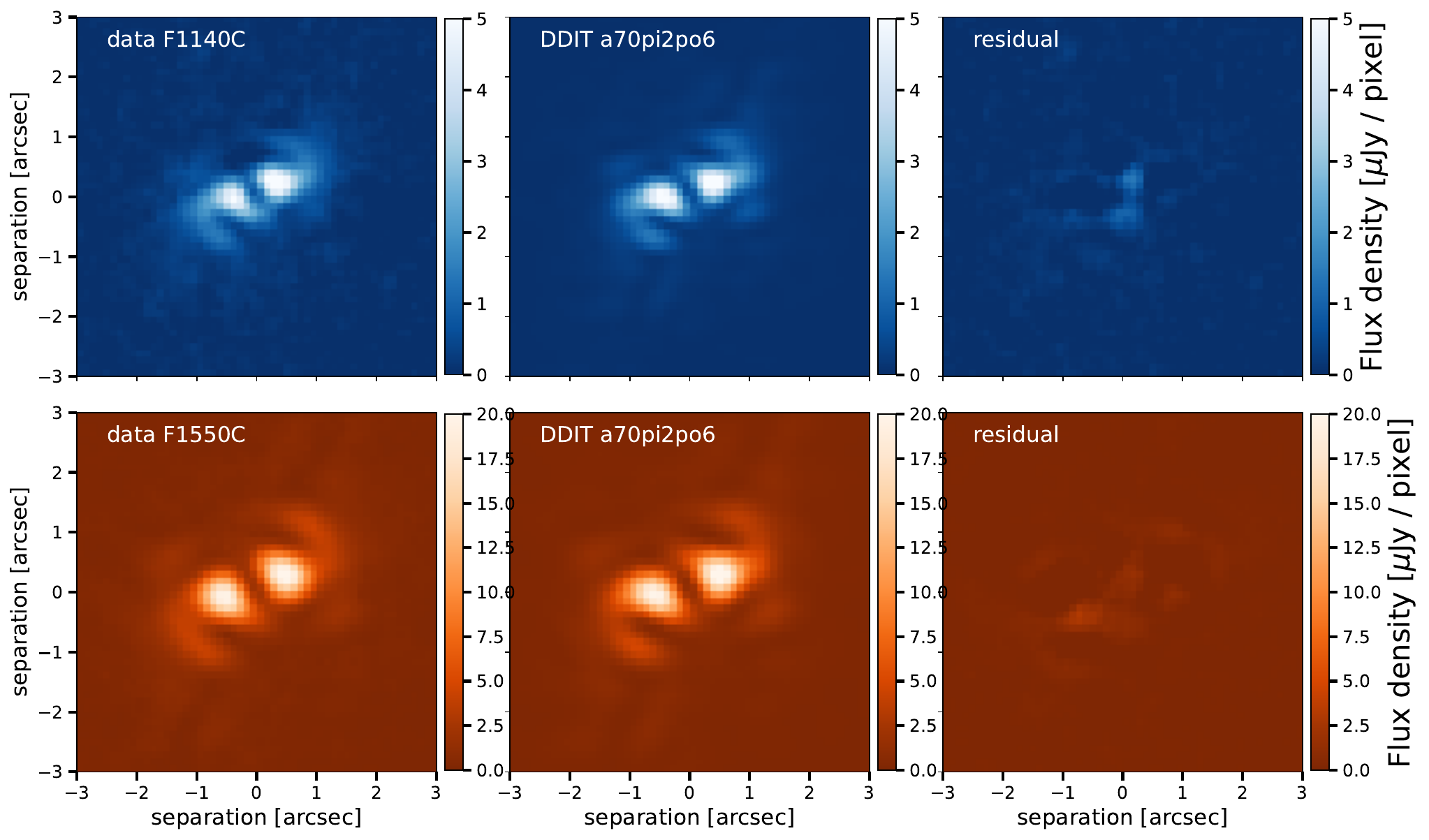}}
	   	\caption{Comparison of the observed coronagraphic images (left) and the 
	   	best simulated images (center). The residuals of the 
	   	difference are plotted at the right. }
	   	\label{fig:F4}
   	\end{figure}

In Table \ref{tab:Table3} we list the selected combinations of parameters that
produced the lowest value of $\chi_{\nu}^2$ ($\nu$ standing for the degree of freedom) 
for $\lambda$ = 15.5 $\mic$ and 11.4 $\mic$, sorted from lowest to highest $\chi_{\nu}^2$. 
The $\chi_{\nu}^2$ is calculated in an elliptical aperture encompassing the disk, of 1.8$''$ 
the major axis, and containing 593 pixels, which cannot be considered independent since 
the angular resolution is 3.2 and 4.4 pixels for F1140C and F1550C, respectively. 
Therefore, we took into account the surface of the PSF in the calculation of the degree of freedom, 
as well as the number of parameters in the geometric model ($R_{c}$, $\alpha_{in}$, and $\alpha_{out}$).
The sets that give the lowest $\chi_{\nu}^2$ are not the same for the two wavelengths, 
but the variation in $\chi_{\nu}^2$  is also not very large: $\pm 11 \%$ for the 12 
listed sets at 15.5 $\mic$, and $\pm 3 \%$ for the 8 sets at 11.4 $\mic$. We decided to only retain the 
set of parameters that was common to both wavelengths. For this set, $\chi_{\nu}^2$ is 
only 22$\%$ and 3.5$\%$ higher than the lowest value at 15.5 $\mic$ 
and 11.4 $\mic$, respectively. This set of parameters, indicated in bold in Table \ref{tab:Table2}, 
features $R_{c}$ = 70 au, $\alpha_{in}$ = 2, and $\alpha_{out}$ = -6. Fig. \ref{fig:F4} 
shows the processed images corresponding to this set of parameters.  
We confirmed this result with a criterion that multiplied the $\chi_{\nu}^2$ for each filter. 
The minimum value was obtained for this very same model : $R_{c}$ = 70 au, $\alpha_{in}$ = 2 ,
and $\alpha_{out}$ = -6.   
We consider this set as the best fit below.
\subsection{The structural parameters}
\label{subsec:structural_param}
 We determined how these quantities compared to previous estimates.
\cite{lagrange2016} and   \cite{kalas2015} used $R_{c}$ = 65 au and a distance of 92 pc at 
a time when this distance  was not firmly established. 
This translates into 74 au at 103 pc, the GAIA distance, which is the most reliable estimate we 
can use today. We note that our best value of $R_{c}$ is similar to the value deduced from a 
radiative transfer modeling of the GPI data by \cite{crotts2021}. 
Because our best fit is for $\alpha_{out}$ = -6, indicating an external 
radius that is not so steeply defined, we considered that 
the near-IR and mid-IR determinations of the critical radius agree well, and we conclude that the grains 
that cause the thermal emission are located at the same distance as those that cause the near-IR scattering. 
With $\alpha_{in}$ = 2, we cannot consider the dust grain population to which we are sensitive at mid-IR as concentrated in a narrow ring coincident with the birth ring of planetesimals, but 
it instead extends well toward the center. 
This agrees with  \cite{bailey2014} , who reported a disk extending from 20 to 120 au 
based on the Ks and L' fluxes, while \cite{fehr2022} deduced a  radially 
broad axisymmetric disk with radii between  50--100 au based on millimeter emission. 
This result does not contradict the fact that in scattered
near-IR light, the appearance of the disk  suggests a void. 
The small grains ($a \sim$ 1 \mic) scatter light at near-IR proportionally 
to their mass fraction,  but emit far more strongly in the mid-IR than their mass fraction because 
they are hottest and the flux in the mid-IR depends exponentially on the temperature 
(Wien part of the Planck's function), as illustrated in Fig. \ref{fig:F5}. A relatively low density of small grains may therefore result in a very significant mid-IR flux. This behavior, which is specific to the mid IR range, was well described theoretically in \citet{thebault2019}.

The total flux for the disk provided by DDiT+ at each wavelength was evaluated in two ways, 
using either the real data and taking the attenuation of the coronagraph derived from 
the total flux ratio of the noncoronagraphic to coronagraphic synthetic images into account (method 1), or in 
the noncoronagraphic synthetic image given the intensity scaling to match the coronagraphic data 
and its corresponding model in the $\chi^2$ process (method 2). For the best model (a70pi2po6), 
we measured a coronagraphic attenuation of 3.0 and 2.5 for F1140C and F1550C, respectively. 
We calculated the mean and the standard deviation of the flux obtained by the model 
in Tab. \ref{tab:Table3} and for the two flux extraction methods. 
The values we retained in the following  are those corresponding to an average of the best 
models: 1223 $\pm$ 20\,µJy and 7230 $\pm$470\,µJy at F1140C and F1550C, respectively. 

\subsection{Flux ratio and grain size}
\label{subsec:flux_ratio}
After we established the structure of the disk, we 
reproduced the observed flux ratios 
$R_f = F_{15.5} / F_{11.4}$ = $5.91\pm 0.50$  
by adjusting the range of grain sizes and the nature of grains. 
The bins of the size follow a geometrical progression of reason 1.38,
starting from $a_{min}$. 

\begin{figure}
	   	\centering
	   	\resizebox{\hsize}{!}{\includegraphics{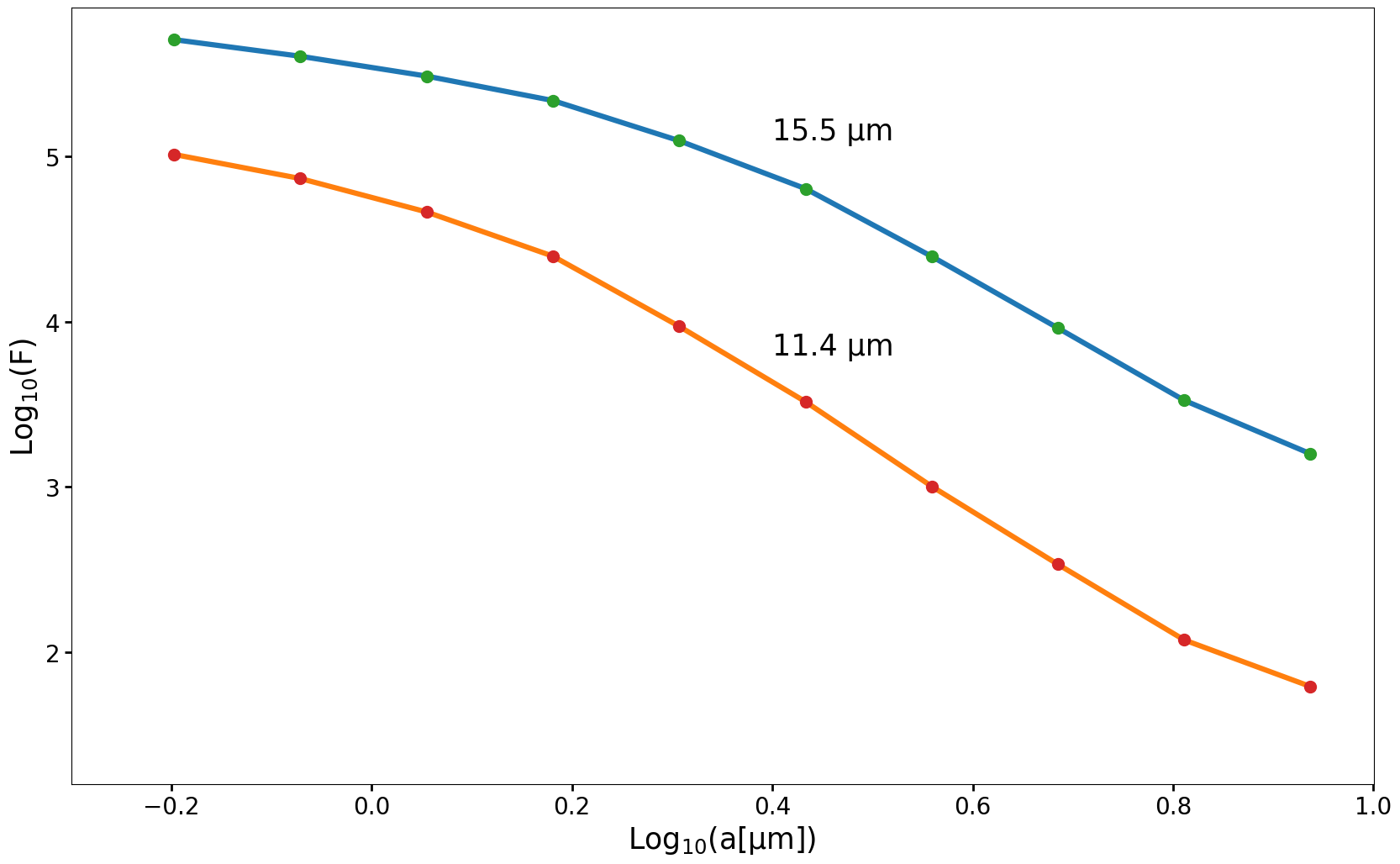}}
	   	\caption{Disk fluxes (in arbitrary units) at 15.5\mic  (red) and 11.4\mic (green)  emitted by each  bin of grain size, as  predicted by DDiT+.  Only the case of silicate grains is shown 
	   	here. }
	   	\label{fig:F5}
   	\end{figure}

\begin{figure}
	   	\centering
	   	\resizebox{\hsize}{!}{\includegraphics{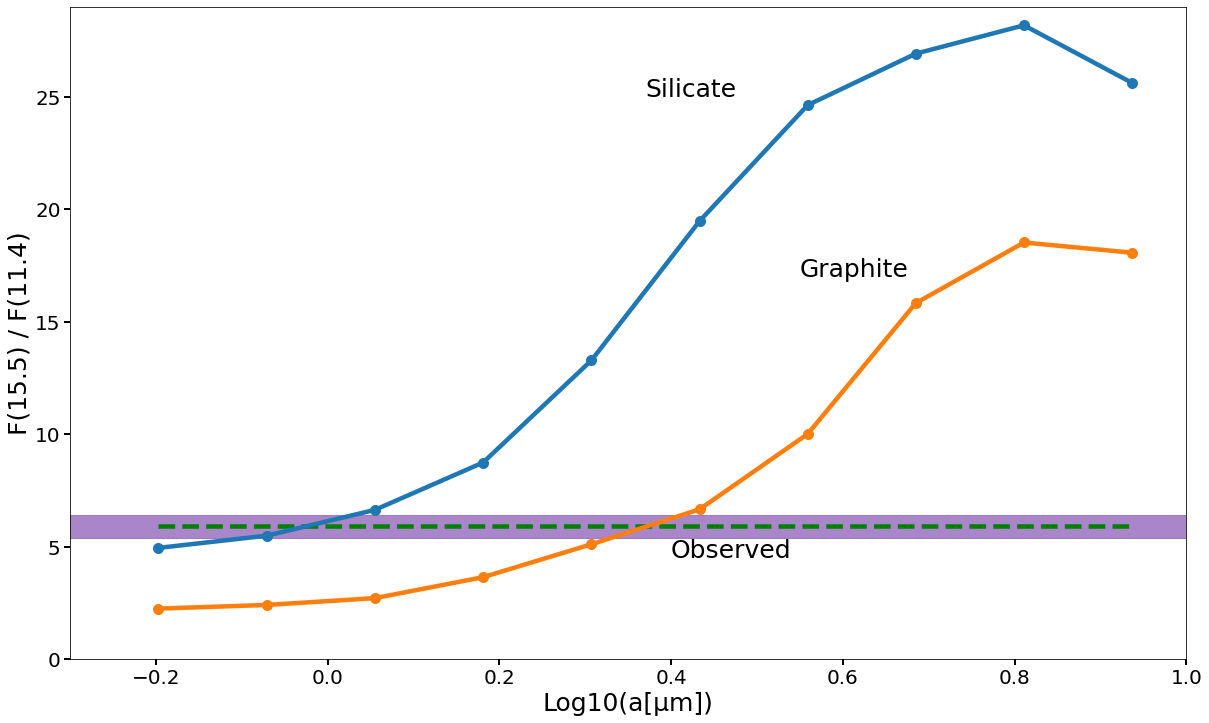}}
	   	\caption{Ratio $R_f = F_{15.5} / F_{11.4}$ vs. grain size for a single 
        grain size, as given by the DDiT+ model. In blue, we show the case of
        silicate grains, and in orange, we show the case of graphite grains. The
        range of ratios deduced from observations is indicated by the 
        shaded purple 
        rectangle. }
	   	\label{fig:F6}
   	\end{figure}

\begin{figure}
	   	\centering
	   	\resizebox{\hsize}{!}{\includegraphics{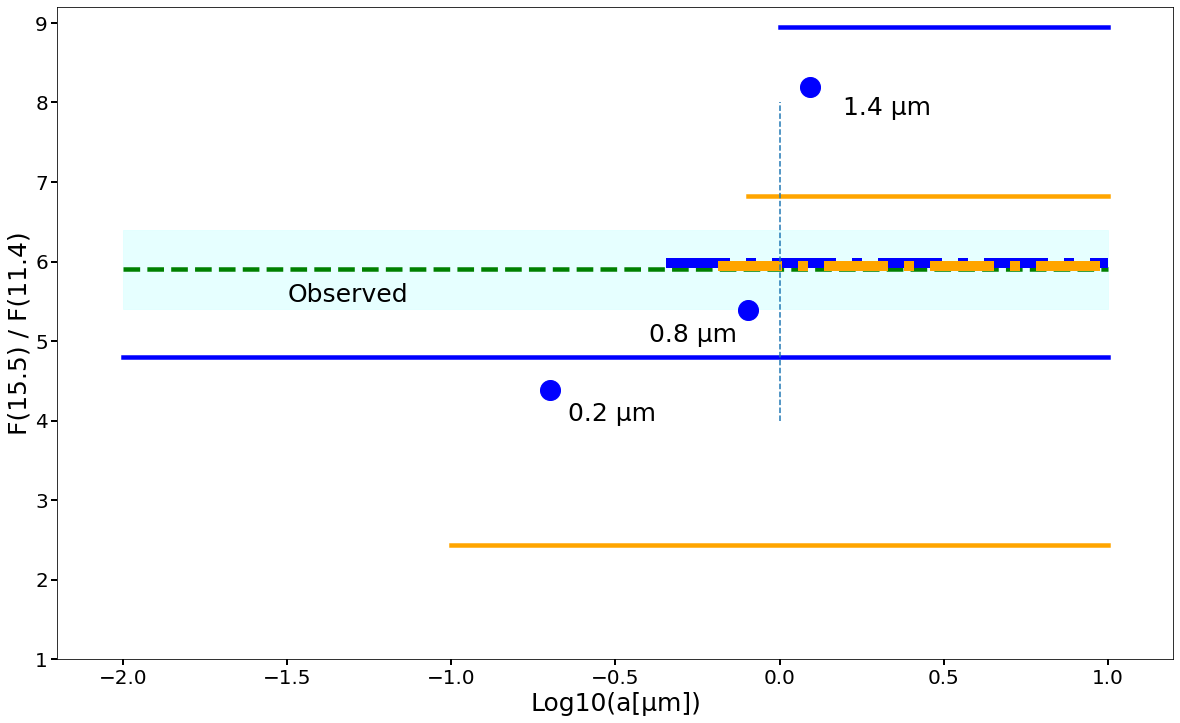}}
	   	\caption{ 
	   	Dependence of the ratio $R_f = F_{15.5} / F_{11.4}$ on the grain size 
	   	distribution. The ratio $R_f$ is deduced from the model after integration 
	   	of the flux in the considered size range. 
	   	$R_f$  is plotted for a selection of different size ranges, all following a distribution in $a^{-3.5}$. 
     Each segment depicts a range that is plotted in blue for silicate grains and in orange for graphite grains. 
     The  dots correspond to single sizes of silicate grains. 
	 The two retained ranges are indicated by thicker dash-dot segments. 
     The  range of ratios deduced from observations is indicated by
        shaded purple.}‹ 
        rectangle. 
	   	\label{fig:F7}
   	\end{figure}

Because the mass fraction associated with each bin 
of size increases from $a_{min}$ to  $a_{max}$ while the corresponding 
fluxes decrease at both wavelengths (Fig. \ref{fig:F5}),  it is not 
intuitive to predict the range that fits the 
observed ratio best. 
For instance, we plot in Fig. \ref{fig:F6} the ratio $R_f$ versus 
the grain size for a population of grains with a unique size. From 
the analysis of this figure, we could be tempted to conclude that the 
maximum size should not be much larger than 1\,$\mic$ for silicate
and 2.5\,$\mic$for graphite, because larger grains, 
when considered alone, produce much higher values of the ratio $R_f$. 
This is not the case, however, and  we show in Fig. \ref{fig:F7} various attempts to 
reproduce the ratio with different choices of $a_{min}$ and  $a_{max}$. 
This figure  indicates that the range should include grains of size $
\sim 1 - 2\,\mic$, but a good fit of the observed ratio is 
reached for silicate grains in the range 0.45 -- 10\,$\mic$ and for graphite
grains in the range  0.65 -- 10 $\mic$, as shown in Fig. \ref{fig:F7}
where these ranges are indicated by thicker segments. 
A rather wide range of sizes is needed to reproduce the 
flux ratio, as expected, 
because in the standard picture of a debris disk, a continuous collisional 
cascade starts at large (unseen) parent bodies and extends to micron-sized 
dust that is blown out by radiation pressure (e.g., \cite{thebault2007}).
 The minimum silicate grain size 
of 0.45\,$\mic$ appears to be consistent with the value that is generally
assumed based on the spectral analysis of the mid-infrared silicate feature 
(\citealt{mittal2015}). Moreover, this value is close to 
the theoretical threshold in radius corresponding
to blowout (see the discussion in \ref{subsec:structDisk}). 
The polarizability curve may also indicate a submicron 
minimum grain size, such as in the case of AU Mic  
(\citealt{graham2007}).  In the following, we consider  two sets of 
parameters that give a reasonable fit to the 
observations. Both have in common  
$\alpha_{in}$ = 2, $\alpha_{out}$ = -6, $R_{c}$ = 70\,au, for silicate 
grains, $a_{min}$ = 0.45\,$\mic$ and $a_{max}$ = 10\,$\mic$, 
and for graphite grains, $a_{min}$ = 0.65\,$\mic$ and $a_{max}$ = 10\,$\mic$. 
We base the remaining discussion on these two sets. 

\section{Discussion} 
\label{sec:discuss}
\subsection{Structure of the disk}
\label{subsec:structDisk}

\begin{figure}
	   	\centering
	   	\resizebox{\hsize}{!}{\includegraphics{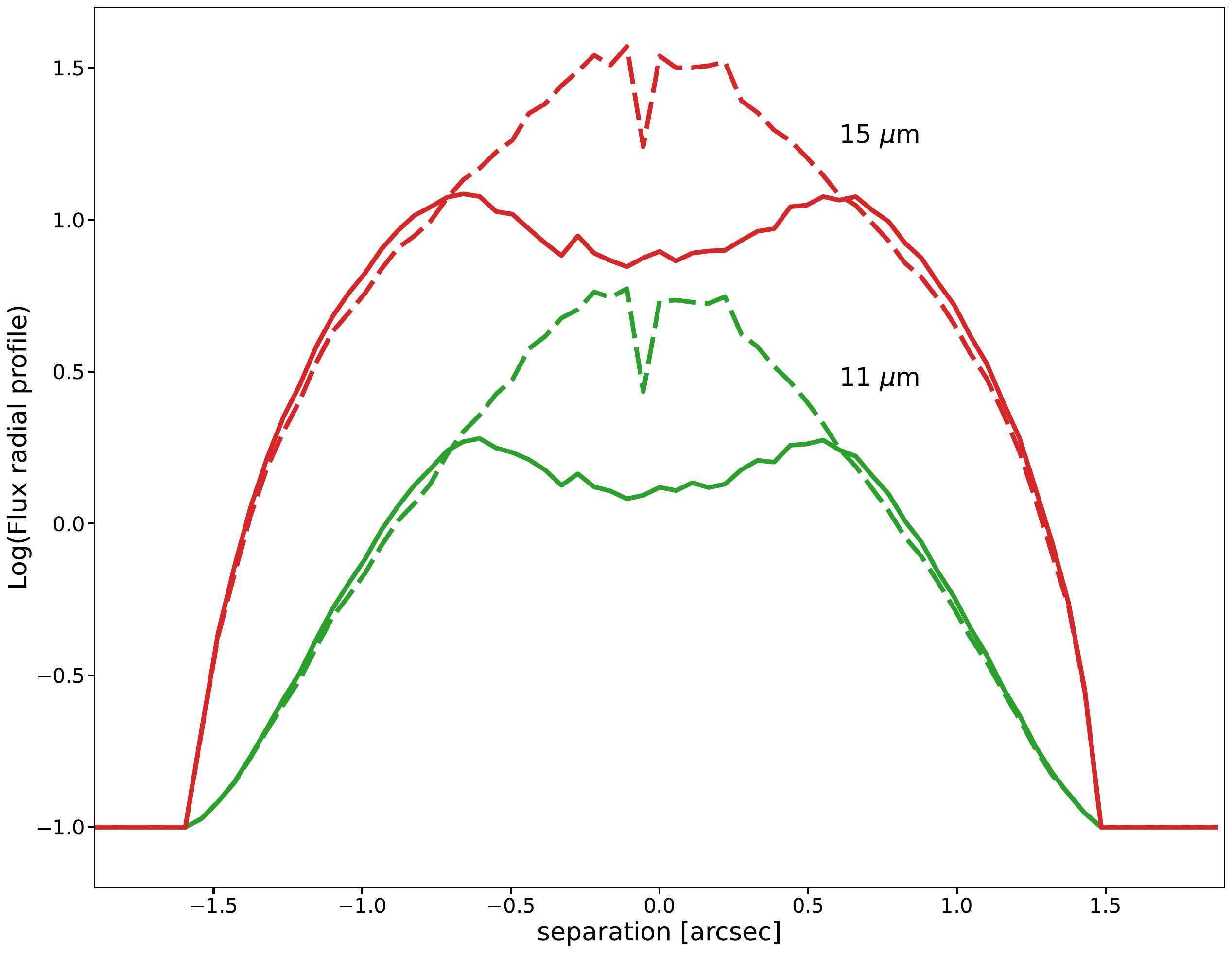}}
	   	\caption{
	   	  Intensity profiles at 11\mic (green) and 15\mic (red) derived from  DDiT+ plotted as a dashed line
	   	in the nominal case ($\alpha_{in}$ = 2, $\alpha_{out}$ = -6) and as a solid line for $\alpha_{in}$ = 6, $\alpha_{out}$ = -6. This illustrates that the forward
	-modeling would be sensitive to the actual dust density profile.}
	   	\label{fig:F8}
   	\end{figure}

In order to reproduce the observed images,
the parameter $\alpha_{in}$ of the model can never be larger than 2, indicating 
an internal disk populated with dust, 
while near-IR observations by \cite{kalas2015} and \cite{lagrange2016} would rather argue for a marked void, possibly generated by an inner planetary systems within a distance of 50\,au. Interestingly,
a power-law index 
of $\alpha_{in}$ = 2 like this is predicted in the case of a purely collisional evolution of
a debris disk, as stated by \cite{Pearce2024}, who proposed an analytic model to investigate 
the effect of a sculpting planet on the radial surface-density profile at the inner disk edge. 
This is often considered as a lack of an internal planet that would otherwise produce a gap or a steep
inner edge, as predicted 
by several models (\citealt{ertel2012, dong2016, ImazBlanco2023}), or observed in numerous cases 
\citep{esposito2020}, for instance, in 
the emblematic system of PDS 70 (\citealt{muller2018, keppler2018, christiaens2024,jang2024, perotti2023}), where the planets carving the cavity are indeed detected. 
To illustrate how MIRI images 
would be sensitive to  differences in profiles 
resulting from a gap, we considered the case of a simulated gap  with $\alpha_{in}$ = 6, 
$\alpha_{out}$ = -6. 
We plot in Fig. \ref{fig:F8} the intensity profiles at 11 and 15 $\mic$ for this case and 
for our nominal case  ($\alpha_{in}$ = 2, $\alpha_{out}$ = -6). 
This gap remains somewhat 
shallow, however, when we consider that the quantity $\sigma_i$  defined by \citep{Pearce2024} to 
quantify the flatness of the inner edge is 0.32 (applying their equation C1 to convert $\alpha_{in}$ = 6 into
$\sigma_i$ ), that is, it is still in the range that is incompatible
with the planet-sculpting only case according to their Fig. 16. Again, we would tend to 
conclude that  there is no massive planet inside the disk. 
Nevertheless, \citep{Pearce2024} listed no fewer than 11 reasons that could cause inner 
edges that are flatter than expected from sculpting planets. 
Finally, we recall that, as discussed above, MIRI does not sense the same range of grain sizes in the thermal IR as in near-IR, which is
entirely dominated by scattering. The two grain populations likely follow a   
different spatial distribution. Therefore, the MIRI observations do not exclude the more or 
less significant presence of dust inside the disk, depending on the grain size.

Another point regarding the structure  of the disk is, as already noted, the slight asymmetry 
between the two lobes. The ESE lobe extends slightly farther  to the ESE, while the WNW 
lobe is slightly brighter and extends slightly more northward. 
This can be observed in particular in the 11.4 \mic image of Fig. \ref{fig:F1} and even more in the right column of 
Fig. \ref{fig:F4}, where the difference between 
the observed image and the image obtained with the model giving the best fit is displayed at each wavelength. The  regions in which the symmetry is  broken are enhanced. They appear to be similar at the two wavelengths.
The relative increase in the brightness is about 20\% $\pm 3.5 \% $.  
This asymmetry is to be compared to the one noted in the near-IR images 
\citep{kalas2015, lagrange2016}, which clearly appears in Fig. \ref{fig:F2}, where the 
near-IR contours describe a  greater vertical width for the  WNW lobe. \cite{nguyen2021} 
described the eastern side of the disk as vertically thin and more extended than the 
western side of the disk, which is vertically thick. This description is consistent with 
what we observe in the mid-IR. In terms of brightness, however, we rather note an anticorrelation 
between the mid-IR and near-IR, the latter exhibiting a brighter SE extension.
This might suggest that as in the Fomalhaut disk \citep{Pan2016}, the disk is eccentric 
and shows an anticorrelated apocenter brightnesses between the near-IR and mid-IR, but some care must be 
exercised because on one hand, the HD106906 disk is seen quasi edge-on,
an unfavorable situation to conclude on its eccentricity, and on the other hand, there is no 
universal rule on this apocenter anticorrelation, as stated by \cite{Lynch2022}, who reported
that at shorter wavelengths the classical pericenter glow effect remains true, 
whereas at longer wavelengths disks can either demonstrate apocenter glow or pericenter glow, 
 depending on the observational resolution. Finally, models showed that at 
 15 \mic, the effect might be undetectable (see Fig. 3 of  \citealt{Pan2016}).
 The difference of the asymmetry between the near-IR and mid-IR can be attributed to the 
 coexistence of several types of grains in the disk and/or a variable dust density, as proposed
 by \cite{Mazoyer2014} (see the discussion of the grain size below).

This asymmetry was proposed to be the result of gravitational perturbations 
by the planet, even though it is very distant \citep{nguyen2021, nesvold2017}. 
This would require either a high eccentricity for the planet (\citealt{nesvold2017} 
propose e = 0.7) or a high relative inclination  between the disk and the planet orbit: for instance \citealt{nguyen2021} proposed 36 or 44\degr, but the error bars are quite large.  
On the other hand, the  N-S  asymmetry can tentatively be interpreted as the trace of a 
vertical warping that might result from a recent close encounter either with a star  \cite{derosa2019} or with a free-floating planet, 
as proposed by \cite{moore2023}. The  binary star at the center may also be at the origin of this asymmetry by introducing some gravitational perturbation in the disk.

The minimum grain size (0.45$\mu$m for silicates) derived from the flux ratio in the two filters agrees with the theoretical estimates of \cite{Kirchschlager2013}, who found
a blowout size of 0.455 \mic for nonporous grains (see their table 3). This is about twice smaller than the  blowout  size (0.85$\mu$m) estimated  by \cite{crotts2021}, however. This situation was also reported in other debris disks and was often associated with a blue color \citep{bhowmik2019, debes2008, augereau2006}, but it seems to contradict the findings of \citet{crotts2021}, who found a bimodal distribution that peaked at 0.91$\mu$m (i.e., consistent with the blowout size) and 3.17$\mu$m (consistent with a neutral color).
However, the blowout size concept does not prevent smaller dust grains from being present in the system because they are continuously produced by collisional cascades in steady state \citep{thebault2019}, which can occur for debris disks with a high fractional luminosity ($f_d>10^{-3}$). We also recall that near-IR polarimetric observations in scattered light presented by \citet{crotts2021}, and these mid-IR observations in the thermal regime, are sensitive to different grains sizes and properties and also to different regions of the disk. For instance, as demonstrated by the value we obtained for the inner slope of the surface density, MIRI sees a population of dust grains inside the birth ring of planetesimals because of their high temperature, and hence, a significant mid-IR emission. The values of $\alpha_{in}$ and $a_{min}$ are likely linked and consistent, as discussed by \citep{thebault2019}.

\subsection{Disk mass}
\label{sec:diskmass}
The range of sizes we selected by fitting the ratio $F_{15.5} / F_{11.4}$ does not exclude the existence of larger grains in the disk: They would indeed not contribute significantly to 
the mid-IR emission, and thus, their contribution is not constrained by the MIRI 
measurements. 
However, using the millimeter fluxes measured with ALMA, we propose estimating 
the fraction and range of these larger grains below and then the total mass of the disk. This is an important 
quantity in all formation scenarii of debris disks and their evolution.  

When the best grain size distribution explaining MIRI data was determined, we evaluated 
the corresponding mass of the disk by simply comparing the fluxes predicted by the model 
assuming 1 $\Mearth$ to the observed fluxes. 
We found $M_{disk}^{s}$ = 0.0033 and 0.0051 $\Mearth$ for silicate and
graphite grains, respectively. The subscript {\it s} means that 
this mass only corresponds to  this range of smaller grains. 
This value was compared to other evaluations. \cite{kral2020} used an MCMC
method to fit millimeter observations and deduced a dust mass of 0.054 $\pm$ 
0.07  $\Mearth$, which is higher by a factor  10 to 16 than our estimates, but they considereds
grains with size up to 1 cm, while we used a maximum size of 10 \mic. 
Based on millimeter data as well, \cite{fehr2022} estimated 10 $\Mearth$ , but without 
detailing how they reached this value. They only mentioned the theoretical work 
of \cite{krivov2021}, who clearly considered  sizes as large as planetesimals. 
This explains the huge discrepancy between the two millimeter-based estimates. 
 
To proceed, we estimated the range of sizes that might account for the 1.27 mm 
fluxes observed by \cite{kral2020} by extrapolating our results. 

A first approach was to assume that the size distribution law is valid up to large grains
with a size of centimeters. In this case, the mass of dust grains with a size up to 1 cm is approximately given
by $M_{disk}^{l} =  M_{disk}^{s} (10000 / 10)^{0.5} $, where the ratio $(10000 / 10) $
corresponds to the ratio of the largest grain sizes (1 cm and  10 \mic) considered for $M_{disk}^{l}$ and  $M_{disk}^{s} $\footnote{because the mass is 
proportional to the integral
$\int_{a_{min}}^{a_{max}} a^3 \, a^-3.5 da = [a_{max}^{.5} -a_{min}^{.5}]$ }, respectively. 
This leads to $M_{disk}^{l}$ $\sim$ 0.11 $\Mearth$ for silicate and 0.16 $\Mearth$
for graphite grains. 
Because of the level of approximation used here, we considered this value  as consistent 
with the mass of 0.054 $\Mearth$  proposed by \cite{kral2020}, which was derived using an entirely 
different approach based on an MCMC estimation. 
The grain density we adopted is $\rho$ = 3.50 kg m$^{-3}$ for silicate. This  
value corresponds to bulk silicate \citep{kimura2003}, the most likely form in a 
debris disk. We used 2.16 kg m$^{-3}$ for graphite grains.
The agreement with \cite{kral2020} would be better with fluffy grains with a density of about 
half the one of bulk material. For instance, \cite{Olofsson2016} reached a better fit with the measured scattering angle in HD 61005by changing
the porosity of silicate grains. Although we obtained a reasonable consistency between our model 
and the various observational data sets (MIRI, ALMA, and near-IR) for silicate 
grains, a mixed composition of dust
that would include amorphous carbon, carboneous compounds or water ice, for instance, 
in addition to silicate, might also lead to an acceptable agreement. 
    
A second approach to confirm the consistency with ALMA data was to start from the flux at 1.27 mm 
emitted by  grains up to 10 \mic, as provided by our best model,  and to estimate the 
contribution of the whole grain population  up to 1cm, that is,  the population 
considered by \cite{kral2020}.  
To do this,  we considered  bins of an increasing grain size, up to 10000 \mic, and we assumed 
that the flux $F_{1270}(a)$ emitted by each bin characterized by size $a$ was proportional to 
$K_o Q_{abs} a^2 a^{-3.5} \Delta a$, where  $\Delta a$ is the range of bin sizes, 
and $ K_o = \Sigma_{grains} B_{1270}(T_{grain})$  
is the sum on all grains of size $a$ in the disk. 

Because  the optical grain properties we used \citep{laor1993} are 
limited in size to a$_{max}$ = 10 \mic and in wavelengths to $\lambda_{max}$ = 1 mm, we had to extrapolate 
the values of Q$_{abs}$ to size up to 1 cm and  to the wavelength $\lambda_0 $= 1.27 mm. In both cases, simple laws give
excellent approximations of Q$_{abs}$. On one hand, 
there is a precise behavior in $\lambda^{-2}$ for the wavelength dependence, and on the other 
hand, for  grains larger than 10 \mic, Q$_{abs}$ only depends upon the quantity $a / \lambda$, so 
that it is always possibleto compute  Q$_{abs}(a, \lambda_0)$ =  
Q$_{abs}( 10,  a \times \lambda_0 / 10 )$  for any size $a > 10$ \mic  (see Appendix A for more details).

\begin{figure}
	   	\centering
	   	\resizebox{\hsize}{!}{\includegraphics{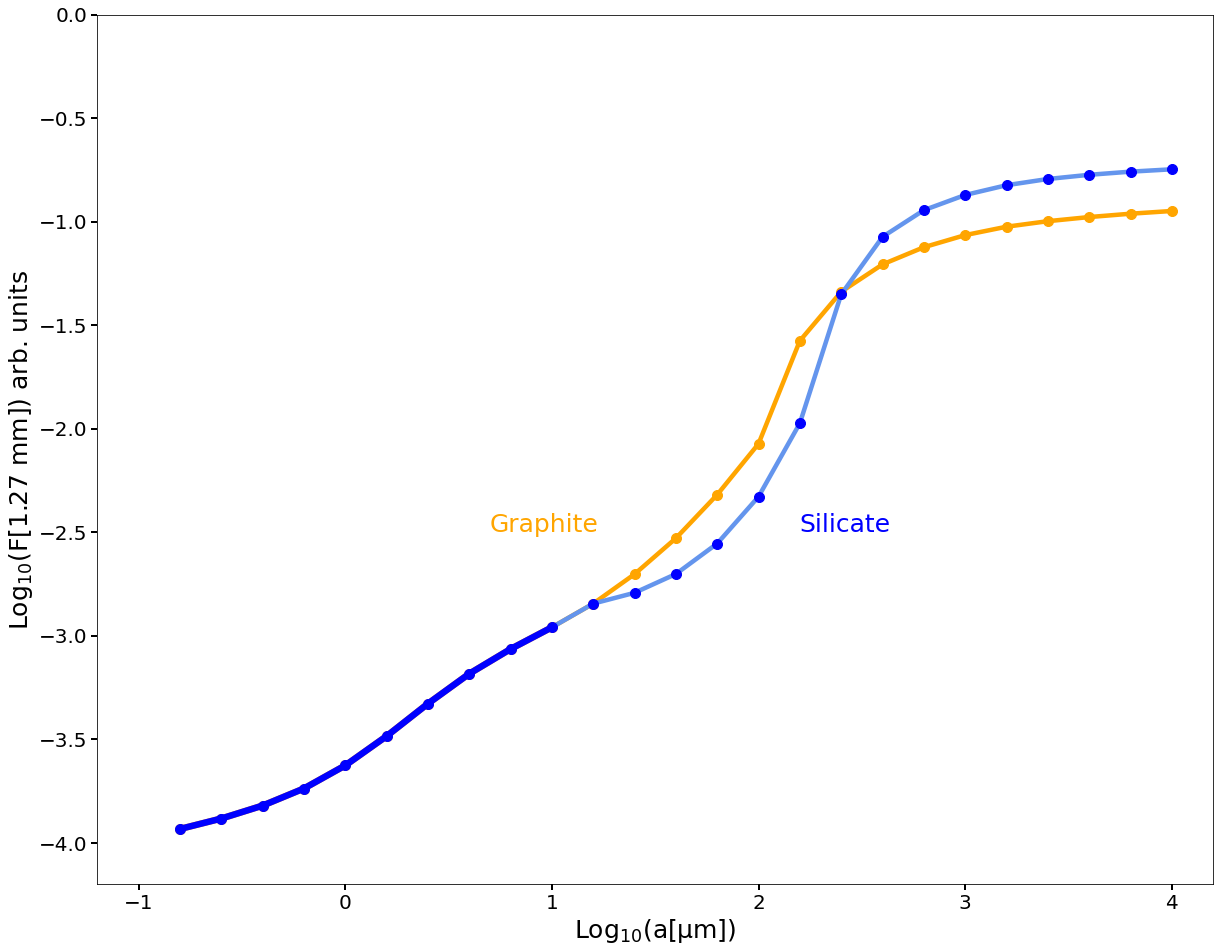}}
	   	\caption{Theoretical estimation  of the flux $F_{1270}$ at 1.27 mm  emitted by the 
	   	population of grains (silicate shown in blue, and graphite shown in orange) up to a given size 
     $a$ (see text). 
     The plotted quantity is proportional to the actual physical quantity. 
	   	The first thicker part of the curve covers the range of sizes (< 10 \mic) for which our
	   	model provides absolute values of the flux.     }
	   	\label{fig:F9}
   	\end{figure}

We then obtained the dependence on $a$ of $F_{1270}(a)$, the flux emitted by each bin, 
and computed the theoretical integrated flux at 1.27 mm emitted by the population of all grains 
up to a given size $a$, as plotted in Fig. \ref{fig:F9}. 
We note that for the largest grains (> 1mm), $F_{1270}(a)$ reaches a plateau that justifies the 
hypothesis that most  of the flux measured by ALMA arises from grains in the 1 - 10 mm range.  
Using this plot, we then estimated the ratio of the flux predicted at 1.27 mm to the flux that is 
solely due to grains of size below 10 \mic: We find $F_{1270}(a <  10 \mic) / F_{1270}(a < 1cm) 
$ = 0.0031 for silicate grains and 0.0127 for graphite grains. 
On the other hand, our best-fit model gives $F_{1270}(a < 10 \mic)$ =  3.73 µJy (17.8 µJy) 
for silicate (graphite) grains, which leads to a predicted flux from the whole population of grains 
$F_{1270}(a < 1cm)$ = 1.20 mJy ( 1.40mJy). Based on the high degree of approximation 
made here, this quantity can also to be considered consistent with the measured ALMA flux of 0.350 mJy 
mentioned by \cite{kral2020}.  
We conclude that our model of the disk, the MIRI measurements, 
and the ALMA observations agree well in general.  
We note that the early estimates of the dust mass in HD 106906, before the detection of the extended 
disk in scattered light, were all  lower that the estimate we propose here. 
For instance, \cite{chen2005} found a lower bound of $1.2 \, 10^{-4} \Mearth $, but 
this was based on simple hypotheses, such as a perfect blackbody emission (while Q$_{abs} 
\approx 0.1 $ at 40 \mic the typical wavelength  of emission of 70 K grains), a grain size of $
\approx 1 \mic$, and a disk with $R_{c}$ = 10 au. 
The similar estimate by \cite{jang-condell2015},  $1.4\, 10^{-4} \Mearth $, also relied 
on some currently unlikely hypotheses : $R_{c}$ = 12 au, and  $T_{gr}$ = 108 K. 
Finally, \cite{mittal2015} proposed a somewhat higher mass of $2.3\, 10^{-3} \Mearth $, but again 
assuming a very compact disk with a radius of 8 au and a rather high value of the exponent in the 
power-law size distribution ($q$ = 4.16).

\subsection{Dust temperature}

\begin{figure}
	   	\centering
	   	\includegraphics[width=9cm]{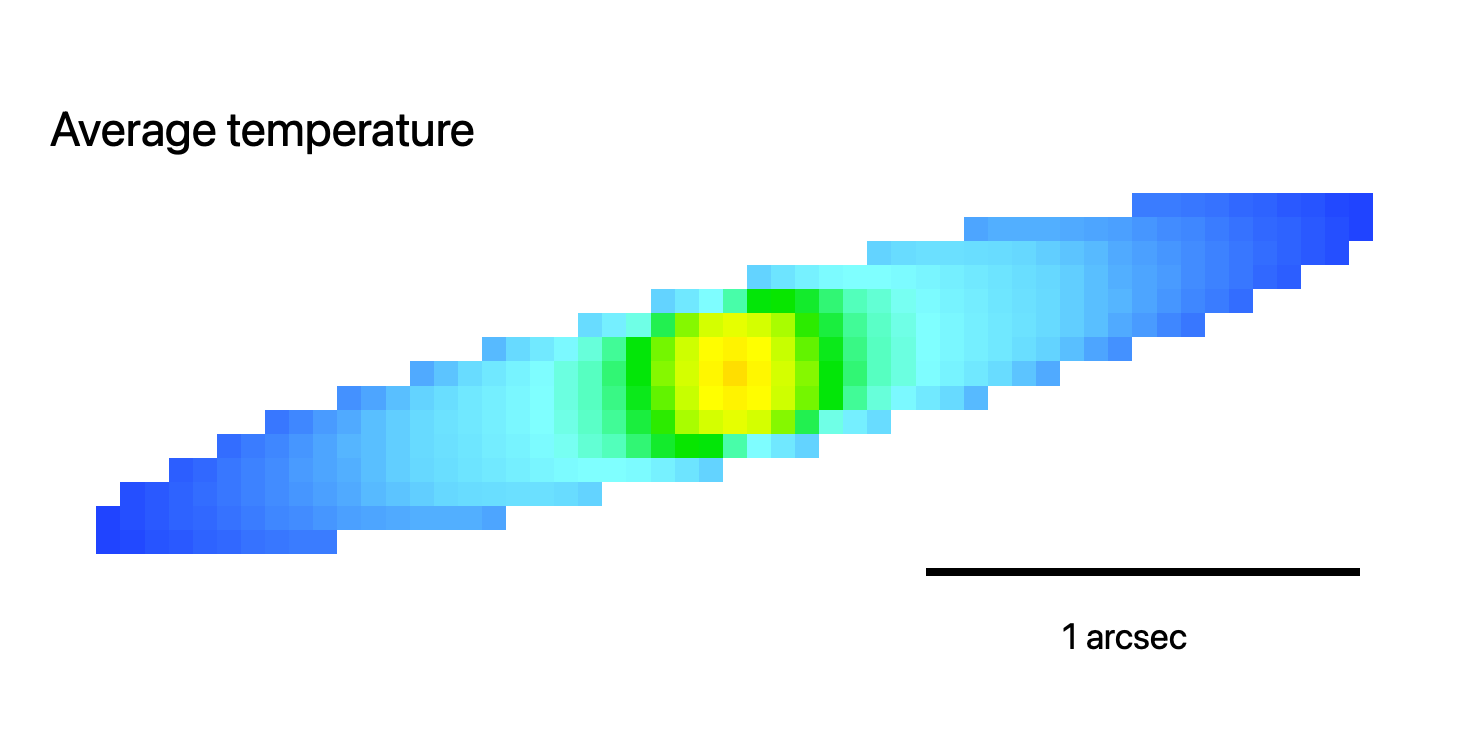}
	   	\includegraphics[width=9cm]{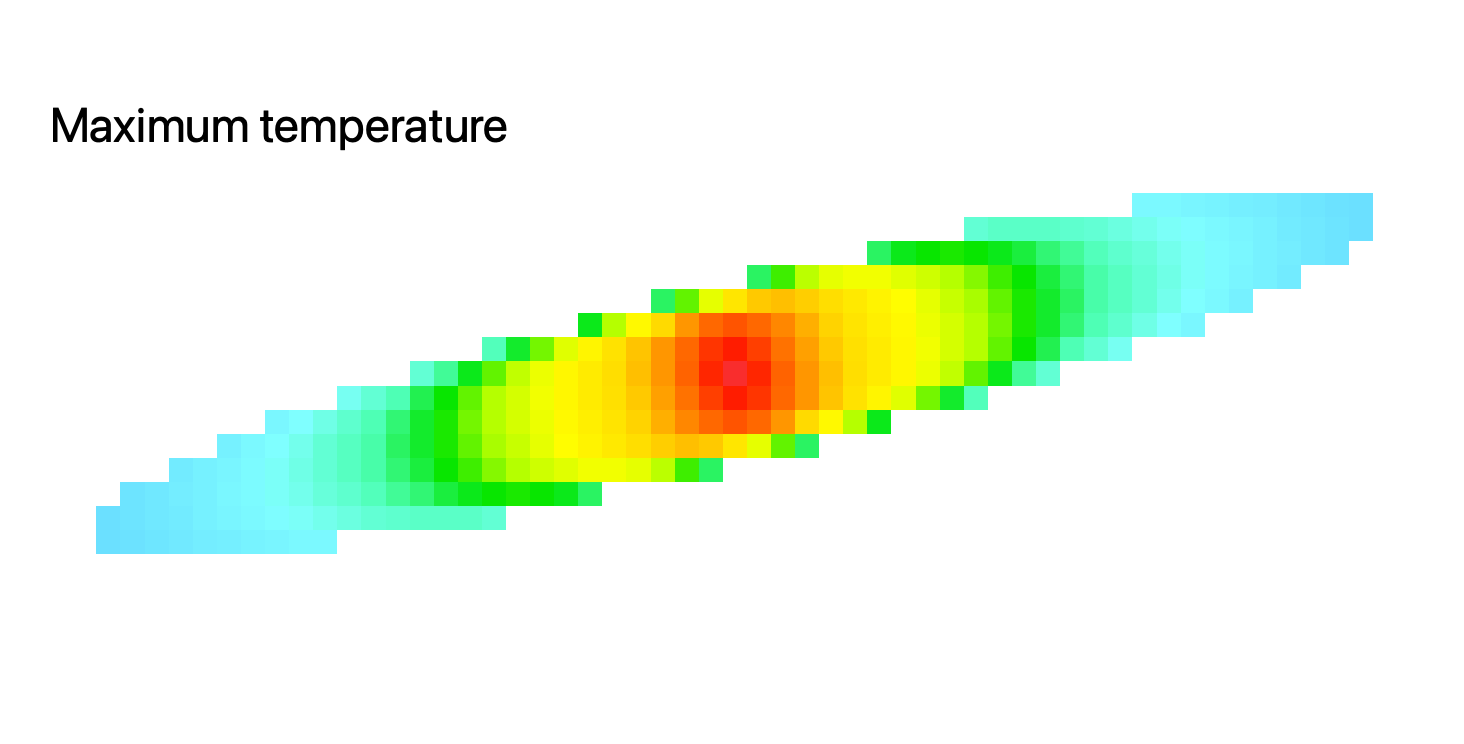}
	   	\includegraphics[width=9cm]{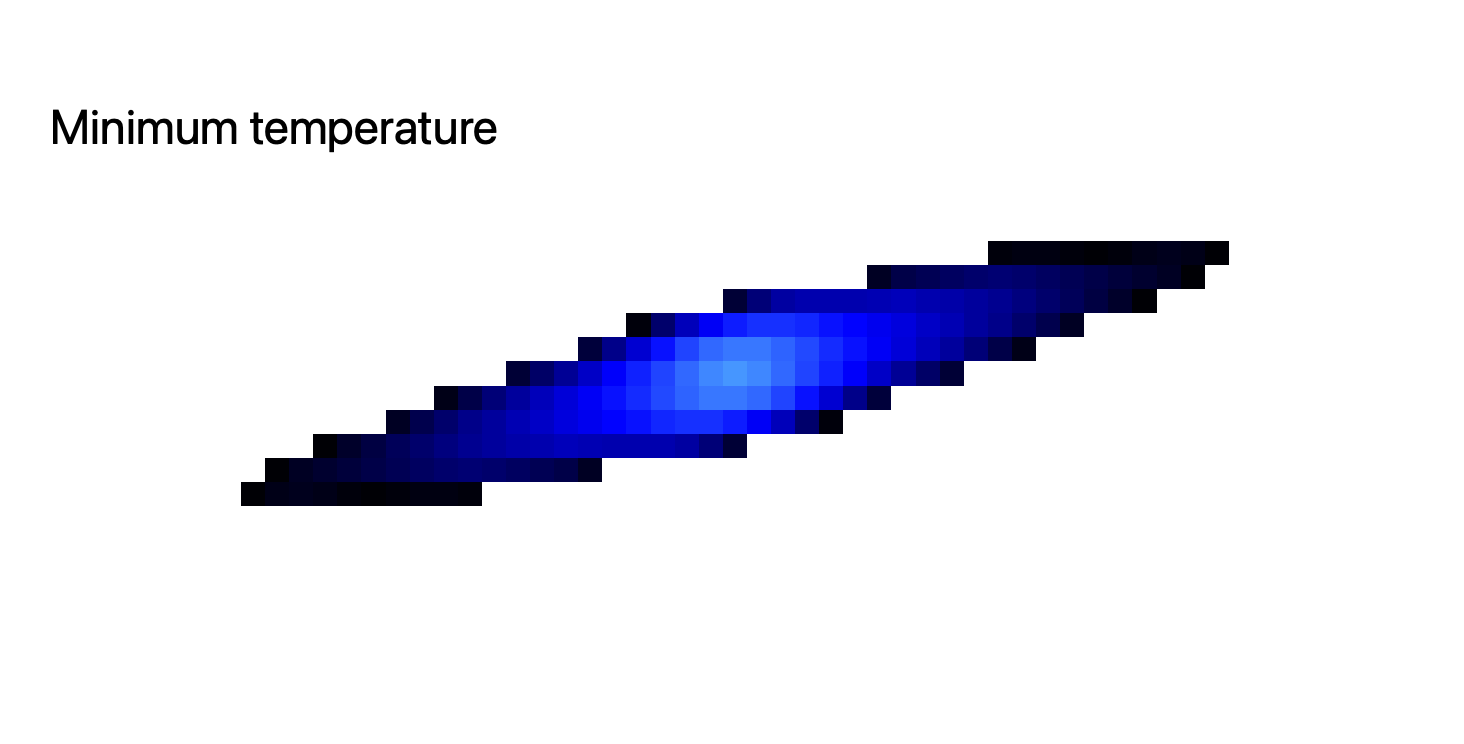}
	   	\includegraphics[width=9cm]{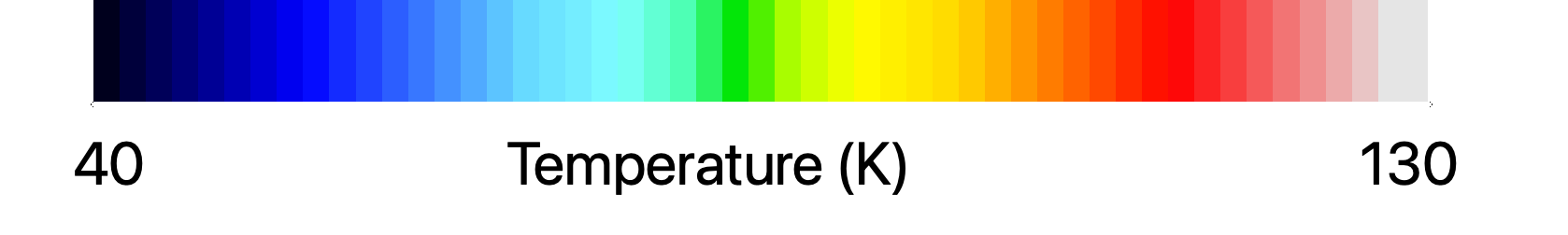}
	   	\caption{From top to bottom: Maps of the average, the maximum, and the minimum dust grain
	   	temperature along each line of sight for the nominal set of parameters with silicate grains.  The color-scale covers
	   	the 	range 40 - 130 K, as indicated at the bottom.}
	   	\label{fig:F10}
   	\end{figure}

The code DDiT+ computed the equilibrium temperature of each grain in the disk and produced
a map of the average temperature along the line of sight. Fig. \ref{fig:F10} displays this 
map for the nominal set of parameters and silicate grains, and the maps of the maximum 
and minimum temperature along the line of sight. We do not show the case of graphite grains, 
but the results are very similar. 
   
The average grain temperature given by the model, considering the whole disk, is 74 K.  
\cite{chen2005} derived a color temperature of 90 K from Spitzer 24 and 70 \mic broadband 
photometry, while \cite{bailey2014}, who also used  data from MIPS-SED spectra up to 100 \mic
 found a fit by a blackbody temperature of 95 K.  
This is  consistent with our result, considering that the  actual range of grain 
temperatures derived from our best model is from 40 K at the edge to 130 K at the center.
Along a given line of sight, the amplitude of the temperature variation is 48 K at the center and 33 
K at the edge. 
The grain temperature is also dependent on the grain size: the smallest grains (a = 0.3 \mic)  
have an average temperature of 93 K, and the largest grains (a = 10 \mic) a temperature of 54 K. 
Since there is essentially no gas in the disk, and thus, no thermalization, except through very 
rare grain-grain collisions, this dispersion in temperature is expected to have consequences on the 
possible formation and segregation of ice mantles, whose condensation temperature lies in the 
range 40 - 130 K, such as methane hydrate (78 K). Water ice  (180 K) or ammonia 
hydrate (131 K) can be present everywhere, if indeed their condensation may occur.   

\subsection{Other planets}
While HD\,106906\,b is well detected and will be studied in a future paper, the MIRI observations also allowed us us to estimate the limit of a detection of closer-in planets. For the sake of simplicity, we assumed only two cases, a planet located inside the disk at 0.5$''$, and another planet farther out at 2.5$''$, both aligned with the disk orientation. The image of each planet was modeled with the same numerical simulation as we used to model the disk in order to account for the coronagraph attenuation. The limits of detection were derived with the injections of simulated planets, from which we measured a star-to-planet contrast of 500 and 7000 at 0.5$''$ and 2.5$''$, respectively. This corresponds to $\Delta M_{F1140C}=6.7$ and 9.6 magnitude. For the evolutionary and atmosphere model \texttt{ATMO2020} \citep{phillips2020}
\footnote{available at \url{http://perso.ens-lyon.fr/isabelle.baraffe/ATMO2020/}} these values translate into a mass of 15\,\Mjup at 0.5$''$ and about 2-3\,\Mjup at 2.5$''$. The detection limits from VLT/SPHERE in \citep{lagrange2016} did not use the same evolutionary models. Starting from their contrast measurements coupled with \texttt{ATMO2020}, we obtained 6\,\Mjup and about 2-3\,\Mjup at the same angular separations. Therefore, inside the disk, MIRI is strongly affected by the intensity of the disk and a poorer angular resolution, while both instruments achieve similar sensitivity in terms
of the planet mass at larger separations.

\section{Conclusions}

We reported successful MIRI-JWST coronagraphic observations of the  debris disk around HD 106906 
at 11.4 and 15.5 \mic. The disk, seen quasi edge-on,  is well detected  at the two wavelengths and extends 
in the direction observed in the near-IR, with the difference, however, that it is  clearly more compact. 
There is some indication of a slight asymmetry between the two lobes that approximately matches the 
asymmetry already noted in the near-IR. 

We analyzed the images using a radiative  transfer code based on the open-source code DDiT 
\citep{olofsson2020} , but we added several modules to produce a realistic model of the scattered and  thermal emission: 
the wavelength-dependent optical properties of grains (silicate or graphite) \citep{laor1993}, 
the distribution of grain sizes as $a^{-3.5}$, the true computation of the radiative equilibrium 
temperature, the  realistic stellar spectrum, and the physical value of the returned fluxes. 
The structure of the disk in the model was controlled by a set of parameters
describing the distribution of dust inside and outside a critical radius. In a second step, 
the images produced by the model were processed by a code that realistically simulated  the whole 
optical path of the JWST-MIRI-coronagraph ensemble.  
The resulting fluxes provided by the model first demonstrated
that the scattered light is totally negligible at these wavelengths and  that only the thermal
emission of the dust that is heated by the central binary star must be considered.
Using a grid of 96 models with different sets of parameters for the structure of the disk, 
we showed that the observed structure is well reproduced by a disk model with a critical radius of 70 au, 
a shallow  increasing density inside (index 2), and a steeper decrease outside (index -6). 
The ratio of fluxes at the two wavelengths strongly constrains the  range of grain sizes  
that cause the mid-IR emission to be between 0.45 -- 10 \mic for silicate grains and 
0.65 -- 10 \mic for graphite grains. We note that the minimum size agrees well with the  
values predicted by some models of the blowout phenomenon due to radiative pressure, but
the difference with other estimates based on near-IR may be due to the probable separate spatial distribution of grains according to their size because mid-IR and near-IR are not sensitive to the same grains. 
We then derived a dust mass that causes the mid-IR emission of $\approx$ 0.0033 -- 0.0050  $\Mearth$ 
and of 0.10 -- 0.16 $\Mearth$ when we extrapolated this to the larger grains (up to 1 cm) that cause the  
millimeter emission, assuming that the law $a^{-3.5}$ of the size distribution is still valid at
large sizes. 
We showed that this estimate is well consistent with ALMA observations of the disk in terms
of  mass and millimeter flux. 

Finally, we provided a map of the dust temperature in the disk
that featured an average value of 74 K, but with a wiede range of temperature within the 
whole disk, from 40 to 130 K. 

We note that the two types of grains we considered, silicate or graphite, can provide a  view 
of the structure and emission of the disk that is consistent with the observations.
We cannot argue on the sole basis of MIRI data in favor of one or the other composition, but we clearly favor the model based on a silicate composition. 
We cannot exclude, however, that a mixture of silicate and inclusions of carbonaceous material are also possible, as proposed, for instance, in \cite{Lebreton2012}. Future spectroscopic observations will likely shed light on this question.

As a final remark, we wish to stress that the MIRI-JWST brings important constraints by giving us
access to the 10-20\,\mic range. This is because the thermal emission by dust in this domain
corresponds to the Wien part of Planck's law, with an exponential behavior with respect to 
wavelength, and thus, a great sensitivity to the actual temperature distribution within the disk. 
The drawback clearly is the likewise high sensitivity to uncertainties, and this 
may explain our failure to reach a strictly firm conclusion on the nature of the grains or 
their size distribution.

\begin{acknowledgements}
We wish to thank warmly Johan Olofsson for having put on Github its DDiT code (\cite{olofsson2020}) 
and for his open mind when he was informed of the evolutions that we brought to the code, as explained 
in the text. 

Daniel Rouan and Anthony Boccaletti thank Philippe Thebault for a fruitful discussion on the 
mass of the disk.  

For the purpose of open access, the authors have applied a Creative Commons Attribution (CC BY) licence to the Author Accepted Manuscript version arising from this submission.

John P. Pye acknowledges financial support from the UK Science and Technology Facilities Council, and the UK Space Agency.
\end{acknowledgements}

\bibliographystyle{aa}  
\bibliography{HD106906_disk} 

\begin{thebibliography}{65}
\expandafter\ifx\csname natexlab\endcsname\relax\def\natexlab#1{#1}\fi

\bibitem[{{Absil} {et~al.}(2021){Absil}, {Marion}, {Ertel}, {Defr{\`e}re},
  {Kennedy}, {Romagnolo}, {Le Bouquin}, {Christiaens}, {Milli}, {Bonsor},
  {Olofsson}, {Su}, \& {Augereau}}]{Absil2021}
{Absil}, O., {Marion}, L., {Ertel}, S., {et~al.} 2021, \aap, 651, A45

\bibitem[{{Argyriou} {et~al.}(2023){Argyriou}, {Lage}, {Rieke}, {Gasman},
  {Bouwman}, {Morrison}, {Libralato}, {Dicken}, {Brandl},
  {{\'A}lvarez-M{\'a}rquez}, {Labiano}, {Regan}, \& {Ressler}}]{Argyriou2023}
{Argyriou}, I., {Lage}, C., {Rieke}, G.~H., {et~al.} 2023, \aap, 680, A96

\bibitem[{Augereau \& Beust(2006)}]{augereau2006}
Augereau, J.~C. \& Beust, H. 2006, Astronomy \& Astrophysics, 455, 987

\bibitem[{{Augereau} {et~al.}(1999){Augereau}, {Lagrange}, {Mouillet},
  {Papaloizou}, \& {Grorod}}]{augereau1999}
{Augereau}, J.~C., {Lagrange}, A.~M., {Mouillet}, D., {Papaloizou}, J.~C.~B.,
  \& {Grorod}, P.~A. 1999, \aap, 348, 557

\bibitem[{{Aumann}(1984)}]{aumann1984}
{Aumann}, H.~H. 1984, in Bulletin of the American Astronomical Society,
  Vol.~16, 483

\bibitem[{{Bailey} {et~al.}(2014){Bailey}, {Meshkat}, {Reiter}, {Morzinski},
  {Males}, {Su}, {Hinz}, {Kenworthy}, {Stark}, {Mamajek}, {Briguglio}, {Close},
  {Follette}, {Puglisi}, {Rodigas}, {Weinberger}, \& {Xompero}}]{bailey2014}
{Bailey}, V., {Meshkat}, T., {Reiter}, M., {et~al.} 2014, \apjl, 780, L4

\bibitem[{Bhowmik {et~al.}(2019)Bhowmik, Boccaletti, Thébault, Kral, Mazoyer,
  Milli, Maire, Holstein, Augereau, Baudoz, Feldt, Galicher, Henning, Lagrange,
  Olofsson, Pantin, \& Perrot}]{bhowmik2019}
Bhowmik, T., Boccaletti, A., Thébault, P., {et~al.} 2019, Astronomy \&
  Astrophysics, 630, A85

\bibitem[{{Boccaletti} {et~al.}(2022){Boccaletti}, {Cossou}, {Baudoz},
  {Lagage}, {Dicken}, {Glasse}, {Hines}, {Aguilar}, {Detre}, {Nickson},
  {Noriega-Crespo}, {G{\'a}sp{\'a}r}, {Labiano}, {Stark}, {Rouan}, {Reess},
  {Wright}, {Rieke}, {Garcia Marin}, {Lajoie}, {Girard}, {Perrin}, {Soummer},
  \& {Pueyo}}]{boccaletti2022}
{Boccaletti}, A., {Cossou}, C., {Baudoz}, P., {et~al.} 2022, \aap, 667, A165

\bibitem[{{Boccaletti} {et~al.}(2015){Boccaletti}, {Lagage}, {Baudoz},
  {Beichman}, {Bouchet}, {Cavarroc}, {Dubreuil}, {Glasse}, {Glauser}, {Hines},
  {Lajoie}, {Lebreton}, {Perrin}, {Pueyo}, {Reess}, {Rieke}, {Ronayette},
  {Rouan}, {Soummer}, \& {Wright}}]{boccaletti2015}
{Boccaletti}, A., {Lagage}, P.~O., {Baudoz}, P., {et~al.} 2015, \pasp, 127, 633

\bibitem[{{Boccaletti} {et~al.}(2024){Boccaletti}, {M{\^a}lin}, {Baudoz},
  {Tremblin}, {Perrot}, {Rouan}, {Lagage}, {Whiteford}, {Molli{\`e}re},
  {Waters}, {Henning}, {Decin}, {G{\"u}del}, {Vandenbussche}, {Absil},
  {Argyriou}, {Bouwman}, {Cossou}, {Coulais}, {Gastaud}, {Glasse}, {Glauser},
  {Kamp}, {Kendrew}, {Krause}, {Lahuis}, {Mueller}, {Olofsson}, {Patapis},
  {Pye}, {Royer}, {Serabyn}, {Scheithauer}, {Colina}, {van Dishoeck}, {Ostlin},
  {Ray}, \& {Wright}}]{boccaletti2024}
{Boccaletti}, A., {M{\^a}lin}, M., {Baudoz}, P., {et~al.} 2024, \aap, 686, A33

\bibitem[{Bushouse {et~al.}(2025)Bushouse, Eisenhamer, Dencheva, Davies,
  Greenfield, Morrison, Hodge, Simon, Grumm, Droettboom, Slavich, Sosey, Pauly,
  Miller, Jedrzejewski, Hack, Davis, Crawford, Law, Gordon, Regan, Cara,
  MacDonald, Bradley, Shanahan, Jamieson, Teodoro, Williams, Pena-Guerrero,
  Graham, Molter, Brandt, Hayes, Cooper, Clarke, \& Filippazzo}]{bushouse2025}
Bushouse, H., Eisenhamer, J., Dencheva, N., {et~al.} 2025, JWST Calibration
  Pipeline

\bibitem[{{Chen} {et~al.}(2005){Chen}, {Patten}, {Werner}, {Dowell},
  {Stapelfeldt}, {Song}, {Stauffer}, {Blaylock}, {Gordon}, \&
  {Krause}}]{chen2005}
{Chen}, C.~H., {Patten}, B.~M., {Werner}, M.~W., {et~al.} 2005, \apj, 634, 1372

\bibitem[{{Christiaens} {et~al.}(2024){Christiaens}, {Samland}, {Henning},
  {Portilla-Revelo}, {Perotti}, {Matthews}, {Absil}, {Decin}, {Kamp},
  {Boccaletti}, {Tabone}, {Marleau}, {van Dishoeck}, {G{\"u}del}, {Lagage},
  {Barrado}, {Caratti o Garatti}, {Glauser}, {Olofsson}, {Ray}, {Scheithauer},
  {Vandenbussche}, {Waters}, {Arabhavi}, {Grant}, {Jang}, {Kanwar},
  {Schreiber}, {Schwarz}, {Temmink}, \& {{\"O}stlin}}]{christiaens2024}
{Christiaens}, V., {Samland}, M., {Henning}, T., {et~al.} 2024, \aap, 685, L1

\bibitem[{{Crotts} {et~al.}(2021){Crotts}, {Matthews}, {Esposito},
  {Duch{\^e}ne}, {Kalas}, {Chen}, {Arriaga}, {Millar-Blanchaer}, {Debes},
  {Draper}, {Fitzgerald}, {Hom}, {MacGregor}, {Mazoyer}, {Patience}, {Rice},
  {Weinberger}, {Wilner}, \& {Wolff}}]{crotts2021}
{Crotts}, K.~A., {Matthews}, B.~C., {Esposito}, T.~M., {et~al.} 2021, \apj,
  915, 58

\bibitem[{{De Rosa} \& {Kalas}(2019)}]{derosa2019}
{De Rosa}, R.~J. \& {Kalas}, P. 2019, \aj, 157, 125

\bibitem[{Debes {et~al.}(2008)Debes, Weinberger, \& Song}]{debes2008}
Debes, J.~H., Weinberger, A.~J., \& Song, I. 2008, The Astrophysical Journal,
  684, L41

\bibitem[{{Dohnanyi}(1968)}]{dohnanyi1968}
{Dohnanyi}, J.~S. 1968, in Physics and Dynamics of Meteors, ed. L.~{Kresak} \&
  P.~M. {Millman}, Vol.~33, 486

\bibitem[{{Dong} \& {Dawson}(2016)}]{dong2016}
{Dong}, R. \& {Dawson}, R. 2016, \apj, 825, 77

\bibitem[{{Ertel} {et~al.}(2012){Ertel}, {Wolf}, \& {Rodmann}}]{ertel2012}
{Ertel}, S., {Wolf}, S., \& {Rodmann}, J. 2012, \aap, 544, A61

\bibitem[{{Esposito} {et~al.}(2020){Esposito}, {Kalas}, {Fitzgerald},
  {Millar-Blanchaer}, {Duch{\^e}ne}, {Patience}, {Hom}, {Perrin}, {De Rosa},
  {Chiang}, {Czekala}, {Macintosh}, {Graham}, {Ansdell}, {Arriaga}, {Bruzzone},
  {Bulger}, {Chen}, {Cotten}, {Dong}, {Draper}, {Follette}, {Hung}, {Lopez},
  {Matthews}, {Mazoyer}, {Metchev}, {Rameau}, {Ren}, {Rice}, {Song}, {Stahl},
  {Wang}, {Wolff}, {Zuckerman}, {Ammons}, {Bailey}, {Barman}, {Chilcote},
  {Doyon}, {Gerard}, {Goodsell}, {Greenbaum}, {Hibon}, {Hinkley}, {Ingraham},
  {Konopacky}, {Maire}, {Marchis}, {Marley}, {Marois}, {Nielsen},
  {Oppenheimer}, {Palmer}, {Poyneer}, {Pueyo}, {Rajan}, {Rantakyr{\"o}},
  {Ruffio}, {Savransky}, {Schneider}, {Sivaramakrishnan}, {Soummer}, {Thomas},
  \& {Ward-Duong}}]{esposito2020}
{Esposito}, T.~M., {Kalas}, P., {Fitzgerald}, M.~P., {et~al.} 2020, \aj, 160,
  24

\bibitem[{{Farhat} {et~al.}(2023){Farhat}, {Sefilian}, \& {Touma}}]{farhat2023}
{Farhat}, M.~A., {Sefilian}, A.~A., \& {Touma}, J.~R. 2023, \mnras, 521, 2067

\bibitem[{{Fehr} {et~al.}(2022){Fehr}, {Hughes}, {Dawson}, {Marino},
  {Ackelsberg}, {Kittling}, {Flaherty}, {Nesvold}, {Carpenter}, {Andrews},
  {Matthews}, {Crotts}, \& {Kalas}}]{fehr2022}
{Fehr}, A.~J., {Hughes}, A.~M., {Dawson}, R.~I., {et~al.} 2022, \apj, 939, 56

\bibitem[{{Gaia Collaboration} {et~al.}(2023){Gaia Collaboration}, {Vallenari},
  {Brown}, {Prusti}, {de Bruijne}, {Arenou}, {Babusiaux}, {Biermann},
  {Creevey}, {Ducourant}, {Evans}, {Eyer}, {Guerra}, {Hutton}, {Jordi},
  {Klioner}, {Lammers}, {Lindegren}, {Luri}, {Mignard}, {Panem}, {Pourbaix},
  {Randich}, {Sartoretti}, {Soubiran}, {Tanga}, {Walton}, {Bailer-Jones},
  {Bastian}, {Drimmel}, {Jansen}, {Katz}, {Lattanzi}, {van Leeuwen}, {Bakker},
  {Cacciari}, {Casta{\~n}eda}, {De Angeli}, {Fabricius}, {Fouesneau},
  {Fr{\'e}mat}, {Galluccio}, {Guerrier}, {Heiter}, {Masana}, {Messineo},
  {Mowlavi}, {Nicolas}, {Nienartowicz}, {Pailler}, {Panuzzo}, {Riclet}, {Roux},
  {Seabroke}, {Sordo}, {Th{\'e}venin}, {Gracia-Abril}, {Portell}, {Teyssier},
  {Altmann}, {Andrae}, {Audard}, {Bellas-Velidis}, {Benson}, {Berthier},
  {Blomme}, {Burgess}, {Busonero}, {Busso}, {C{\'a}novas}, {Carry}, {Cellino},
  {Cheek}, {Clementini}, {Damerdji}, {Davidson}, {de Teodoro}, {Nu{\~n}ez
  Campos}, {Delchambre}, {Dell'Oro}, {Esquej}, {Fern{\'a}ndez-Hern{\'a}ndez},
  {Fraile}, {Garabato}, {Garc{\'\i}a-Lario}, {Gosset}, {Haigron}, {Halbwachs},
  {Hambly}, {Harrison}, {Hern{\'a}ndez}, {Hestroffer}, {Hodgkin}, {Holl},
  {Jan{\ss}en}, {Jevardat de Fombelle}, {Jordan}, {Krone-Martins}, {Lanzafame},
  {L{\"o}ffler}, {Marchal}, {Marrese}, {Moitinho}, {Muinonen}, {Osborne},
  {Pancino}, {Pauwels}, {Recio-Blanco}, {Reyl{\'e}}, {Riello}, {Rimoldini},
  {Roegiers}, {Rybizki}, {Sarro}, {Siopis}, {Smith}, {Sozzetti}, {Utrilla},
  {van Leeuwen}, {Abbas}, {{\'A}brah{\'a}m}, {Abreu Aramburu}, {Aerts},
  {Aguado}, {Ajaj}, {Aldea-Montero}, {Altavilla}, {{\'A}lvarez}, {Alves},
  {Anders}, {Anderson}, {Anglada Varela}, {Antoja}, {Baines}, {Baker},
  {Balaguer-N{\'u}{\~n}ez}, {Balbinot}, {Balog}, {Barache}, {Barbato},
  {Barros}, {Barstow}, {Bartolom{\'e}}, {Bassilana}, {Bauchet}, {Becciani},
  {Bellazzini}, {Berihuete}, {Bernet}, {Bertone}, {Bianchi}, {Binnenfeld},
  {Blanco-Cuaresma}, {Blazere}, {Boch}, {Bombrun}, {Bossini}, {Bouquillon},
  {Bragaglia}, {Bramante}, {Breedt}, {Bressan}, {Brouillet}, {Brugaletta},
  {Bucciarelli}, {Burlacu}, {Butkevich}, {Buzzi}, {Caffau}, {Cancelliere},
  {Cantat-Gaudin}, {Carballo}, {Carlucci}, {Carnerero}, {Carrasco},
  {Casamiquela}, {Castellani}, {Castro-Ginard}, {Chaoul}, {Charlot}, {Chemin},
  {Chiaramida}, {Chiavassa}, {Chornay}, {Comoretto}, {Contursi}, {Cooper},
  {Cornez}, {Cowell}, {Crifo}, {Cropper}, {Crosta}, {Crowley}, {Dafonte},
  {Dapergolas}, {David}, {David}, {de Laverny}, {De Luise}, {De March}, {De
  Ridder}, {de Souza}, {de Torres}, {del Peloso}, {del Pozo}, {Delbo},
  {Delgado}, {Delisle}, {Demouchy}, {Dharmawardena}, {Di Matteo}, {Diakite},
  {Diener}, {Distefano}, {Dolding}, {Edvardsson}, {Enke}, {Fabre}, {Fabrizio},
  {Faigler}, {Fedorets}, {Fernique}, {Fienga}, {Figueras}, {Fournier},
  {Fouron}, {Fragkoudi}, {Gai}, {Garcia-Gutierrez}, {Garcia-Reinaldos},
  {Garc{\'\i}a-Torres}, {Garofalo}, {Gavel}, {Gavras}, {Gerlach}, {Geyer},
  {Giacobbe}, {Gilmore}, {Girona}, {Giuffrida}, {Gomel}, {Gomez},
  {Gonz{\'a}lez-N{\'u}{\~n}ez}, {Gonz{\'a}lez-Santamar{\'\i}a},
  {Gonz{\'a}lez-Vidal}, {Granvik}, {Guillout}, {Guiraud},
  {Guti{\'e}rrez-S{\'a}nchez}, {Guy}, {Hatzidimitriou}, {Hauser}, {Haywood},
  {Helmer}, {Helmi}, {Sarmiento}, {Hidalgo}, {Hilger}, {H{\l}adczuk}, {Hobbs},
  {Holland}, {Huckle}, {Jardine}, {Jasniewicz}, {Jean-Antoine Piccolo},
  {Jim{\'e}nez-Arranz}, {Jorissen}, {Juaristi Campillo}, {Julbe}, {Karbevska},
  {Kervella}, {Khanna}, {Kontizas}, {Kordopatis}, {Korn}, {K{\'o}sp{\'a}l},
  {Kostrzewa-Rutkowska}, {Kruszy{\'n}ska}, {Kun}, {Laizeau}, {Lambert},
  {Lanza}, {Lasne}, {Le Campion}, {Lebreton}, {Lebzelter}, {Leccia}, {Leclerc},
  {Lecoeur-Taibi}, {Liao}, {Licata}, {Lindstr{\o}m}, {Lister}, {Livanou},
  {Lobel}, {Lorca}, {Loup}, {Madrero Pardo}, {Magdaleno Romeo}, {Managau},
  {Mann}, {Manteiga}, {Marchant}, {Marconi}, {Marcos}, {Marcos Santos},
  {Mar{\'\i}n Pina}, {Marinoni}, {Marocco}, {Marshall}, {Martin Polo},
  {Mart{\'\i}n-Fleitas}, {Marton}, {Mary}, {Masip}, {Massari},
  {Mastrobuono-Battisti}, {Mazeh}, {McMillan}, {Messina}, {Michalik}, {Millar},
  {Mints}, {Molina}, {Molinaro}, {Moln{\'a}r}, {Monari}, {Mongui{\'o}},
  {Montegriffo}, {Montero}, {Mor}, {Mora}, {Morbidelli}, {Morel}, {Morris},
  {Muraveva}, {Murphy}, {Musella}, {Nagy}, {Noval}, {Oca{\~n}a}, {Ogden},
  {Ordenovic}, {Osinde}, {Pagani}, {Pagano}, {Palaversa}, {Palicio},
  {Pallas-Quintela}, {Panahi}, {Payne-Wardenaar}, {Pe{\~n}alosa Esteller},
  {Penttil{\"a}}, {Pichon}, {Piersimoni}, {Pineau}, {Plachy}, {Plum}, {Poggio},
  {Pr{\v{s}}a}, {Pulone}, {Racero}, {Ragaini}, {Rainer}, {Raiteri}, {Rambaux},
  {Ramos}, {Ramos-Lerate}, {Re Fiorentin}, {Regibo}, {Richards}, {Rios Diaz},
  {Ripepi}, {Riva}, {Rix}, {Rixon}, {Robichon}, {Robin}, {Robin}, {Roelens},
  {Rogues}, {Rohrbasser}, {Romero-G{\'o}mez}, {Rowell}, {Royer}, {Ruz Mieres},
  {Rybicki}, {Sadowski}, {S{\'a}ez N{\'u}{\~n}ez}, {Sagrist{\`a} Sell{\'e}s},
  {Sahlmann}, {Salguero}, {Samaras}, {Sanchez Gimenez}, {Sanna},
  {Santove{\~n}a}, {Sarasso}, {Schultheis}, {Sciacca}, {Segol}, {Segovia},
  {S{\'e}gransan}, {Semeux}, {Shahaf}, {Siddiqui}, {Siebert}, {Siltala},
  {Silvelo}, {Slezak}, {Slezak}, {Smart}, {Snaith}, {Solano}, {Solitro},
  {Souami}, {Souchay}, {Spagna}, {Spina}, {Spoto}, {Steele},
  {Steidelm{\"u}ller}, {Stephenson}, {S{\"u}veges}, {Surdej}, {Szabados},
  {Szegedi-Elek}, {Taris}, {Taylor}, {Teixeira}, {Tolomei}, {Tonello}, {Torra},
  {Torra}, {Torralba Elipe}, {Trabucchi}, {Tsounis}, {Turon}, {Ulla}, {Unger},
  {Vaillant}, {van Dillen}, {van Reeven}, {Vanel}, {Vecchiato}, {Viala},
  {Vicente}, {Voutsinas}, {Weiler}, {Wevers}, {Wyrzykowski}, {Yoldas}, {Yvard},
  {Zhao}, {Zorec}, {Zucker}, \& {Zwitter}}]{gaia2023}
{Gaia Collaboration}, {Vallenari}, A., {Brown}, A.~G.~A., {et~al.} 2023, \aap,
  674, A1

\bibitem[{{Graham} {et~al.}(2007){Graham}, {Kalas}, \& {Matthews}}]{graham2007}
{Graham}, J.~R., {Kalas}, P.~G., \& {Matthews}, B.~C. 2007, \apj, 654, 595

\bibitem[{{Hughes} {et~al.}(2018{\natexlab{a}}){Hughes}, {Duch{\^e}ne}, \&
  {Matthews}}]{hughes2016}
{Hughes}, A.~M., {Duch{\^e}ne}, G., \& {Matthews}, B.~C. 2018{\natexlab{a}},
  \araa, 56, 541

\bibitem[{{Hughes} {et~al.}(2018{\natexlab{b}}){Hughes}, {Duch{\^e}ne}, \&
  {Matthews}}]{hughes2018}
{Hughes}, A.~M., {Duch{\^e}ne}, G., \& {Matthews}, B.~C. 2018{\natexlab{b}},
  \araa, 56, 541

\bibitem[{{Hughes} {et~al.}(2018{\natexlab{c}}){Hughes}, {Duch{\^e}ne}, \&
  {Matthews}}]{Hugues2018}
{Hughes}, A.~M., {Duch{\^e}ne}, G., \& {Matthews}, B.~C. 2018{\natexlab{c}},
  \araa, 56, 541

\bibitem[{{Imaz Blanco} {et~al.}(2023){Imaz Blanco}, {Marino}, {Matr{\`a}},
  {Booth}, {Carpenter}, {Faramaz}, {Henning}, {Hughes}, {Kennedy}, {P{\'e}rez},
  {Ricci}, \& {Wyatt}}]{ImazBlanco2023}
{Imaz Blanco}, A., {Marino}, S., {Matr{\`a}}, L., {et~al.} 2023, \mnras, 522,
  6150

\bibitem[{{Jang} {et~al.}(2024){Jang}, {Waters}, {Kaeufer}, {Tamanai},
  {Perotti}, {Christiaens}, {Kamp}, {Henning}, {Min}, {Arabhavi}, {Barrado},
  {van Dishoeck}, {Gasman}, {Grant}, {G{\"u}del}, {Lagage}, {Lahuis},
  {Schwarz}, {Tabone}, \& {Temmink}}]{jang2024}
{Jang}, H., {Waters}, R., {Kaeufer}, T., {et~al.} 2024, arXiv e-prints,
  arXiv:2408.16367

\bibitem[{{Jang-Condell} {et~al.}(2015){Jang-Condell}, {Chen}, {Mittal},
  {Manoj}, {Watson}, {Lisse}, {Nesvold}, \& {Kuchner}}]{jang-condell2015}
{Jang-Condell}, H., {Chen}, C.~H., {Mittal}, T., {et~al.} 2015, \apj, 808, 167

\bibitem[{{Kalas} {et~al.}(2015){Kalas}, {Rajan}, {Wang}, {Millar-Blanchaer},
  {Duchene}, {Chen}, {Fitzgerald}, {Dong}, {Graham}, {Patience}, {Macintosh},
  {Murray-Clay}, {Matthews}, {Rameau}, {Marois}, {Chilcote}, {De Rosa},
  {Doyon}, {Draper}, {Lawler}, {Ammons}, {Arriaga}, {Bulger}, {Cotten},
  {Follette}, {Goodsell}, {Greenbaum}, {Hibon}, {Hinkley}, {Hung}, {Ingraham},
  {Konapacky}, {Lafreniere}, {Larkin}, {Long}, {Maire}, {Marchis}, {Metchev},
  {Morzinski}, {Nielsen}, {Oppenheimer}, {Perrin}, {Pueyo}, {Rantakyr{\"o}},
  {Ruffio}, {Saddlemyer}, {Savransky}, {Schneider}, {Sivaramakrishnan},
  {Soummer}, {Song}, {Thomas}, {Vasisht}, {Ward-Duong}, {Wiktorowicz}, \&
  {Wolff}}]{kalas2015}
{Kalas}, P.~G., {Rajan}, A., {Wang}, J.~J., {et~al.} 2015, \apj, 814, 32

\bibitem[{Keppler {et~al.}(2018)Keppler, benisty, Müller, Henning, Boekel,
  Cantalloube, Ginski, Holstein, Maire, Pohl, Samland, Avenhaus, Baudino,
  Boccaletti, Boer, Bonnefoy, Chauvin, Desidera, Langlois, Lazzoni, Marleau,
  Mordasini, Pawellek, Stolker, Vigan, Zurlo, Birnstiel, Brandner, Feldt,
  Flock, Girard, Gratton, Hagelberg, Isella, Janson, Juhasz, Kemmer, Kral,
  Lagrange, Launhardt, Matter, Menard, Milli, Mollière, Olofsson, Perez,
  Pinilla, Pinte, Quanz, Schmidt, Udry, Wahhaj, Williams, Buenzli, Cudel,
  Dominik, Galicher, Kasper, Lannier, Mesa, Mouillet, Peretti, Perrot, Salter,
  Sissa, Wildi, Abe, Antichi, Augereau, Baruffolo, Baudoz, Bazzon, Beuzit,
  Blanchard, Brems, Buey, Caprio, Carbillet, Carle, Cascone, Cheetham, Claudi,
  Costille, Delboulbe, Dohlen, Fantinel, Feautrier, Fusco, Giro, Gluck, Gry,
  Hubin, Hugot, Jaquet, Mignant, Llored, Madec, Magnard, Martinez, Maurel,
  Meyer, Möller-Nilsson, Moulin, Mugnier, Origne, Pavlov, Perret, Petit,
  Pragt, Puget, Rabou, Ramos, Rigal, Rochat, Roelfsema, Rousset, Roux,
  Salasnich, Sauvage, Sevin, Soenke, Stadler, Suarez, Turatto, \&
  Weber}]{keppler2018}
Keppler, M., benisty, M., Müller, A., {et~al.} 2018, Astronomy \&
  Astrophysics, 617, A44

\bibitem[{{Kimura} {et~al.}(2003){Kimura}, {Mann}, {Jessberger}, \&
  {Weber}}]{kimura2003}
{Kimura}, H., {Mann}, I., {Jessberger}, E.~K., \& {Weber}, I. 2003, Meteoritics
  and Planetary Science Supplement, 38, 5211

\bibitem[{{Kirchschlager} \& {Wolf}(2013)}]{Kirchschlager2013}
{Kirchschlager}, F. \& {Wolf}, S. 2013, in Protostars and Planets VI Posters

\bibitem[{{Kral} {et~al.}(2020){Kral}, {Matr{\`a}}, {Kennedy}, {Marino}, \&
  {Wyatt}}]{kral2020}
{Kral}, Q., {Matr{\`a}}, L., {Kennedy}, G.~M., {Marino}, S., \& {Wyatt}, M.~C.
  2020, \mnras, 497, 2811

\bibitem[{{Krivov} \& {Wyatt}(2021)}]{krivov2021}
{Krivov}, A.~V. \& {Wyatt}, M.~C. 2021, \mnras, 500, 718

\bibitem[{{Lagrange} {et~al.}(2016){Lagrange}, {Langlois}, {Gratton}, {Maire},
  {Milli}, {Olofsson}, {Vigan}, {Bailey}, {Mesa}, {Chauvin}, {Boccaletti},
  {Galicher}, {Girard}, {Bonnefoy}, {Samland}, {Menard}, {Henning},
  {Kenworthy}, {Thalmann}, {Beust}, {Beuzit}, {Brandner}, {Buenzli},
  {Cheetham}, {Janson}, {le Coroller}, {Lannier}, {Mouillet}, {Peretti},
  {Perrot}, {Salter}, {Sissa}, {Wahhaj}, {Abe}, {Desidera}, {Feldt}, {Madec},
  {Perret}, {Petit}, {Rabou}, {Soenke}, \& {Weber}}]{lagrange2016}
{Lagrange}, A.~M., {Langlois}, M., {Gratton}, R., {et~al.} 2016, \aap, 586, L8

\bibitem[{{Laor} \& {Draine}(1993)}]{laor1993}
{Laor}, A. \& {Draine}, B.~T. 1993, \apj, 402, 441

\bibitem[{{Lebreton} {et~al.}(2012){Lebreton}, {Augereau}, {Thi}, {Roberge},
  {Donaldson}, {Schneider}, {Maddison}, {M{\'e}nard}, {Riviere-Marichalar},
  {Mathews}, {Kamp}, {Pinte}, {Dent}, {Barrado}, {Duch{\^e}ne}, {Gonzalez},
  {Grady}, {Meeus}, {Pantin}, {Williams}, \& {Woitke}}]{Lebreton2012}
{Lebreton}, J., {Augereau}, J.~C., {Thi}, W.~F., {et~al.} 2012, \aap, 539, A17

\bibitem[{{Lynch} \& {Lovell}(2022)}]{Lynch2022}
{Lynch}, E.~M. \& {Lovell}, J.~B. 2022, \mnras, 510, 2538

\bibitem[{{M{\^a}lin} {et~al.}(2024){M{\^a}lin}, {Boccaletti}, {Baudoz},
  {Perrot}, {Rouan}, {-O.}, {Whiteford}, {Molli{\`e}re}, {Waters}, {Henning},
  {Decin}, {G{\"u}del}, {Vadenbussche}, {Absil}, {Argyriou}, {Bouwman},
  {Cossou}, {Coulais}, {Gastaud}, {Glasse}, {Glauser}, {Kamp}, {Kendrew},
  {Krause}, {Lahuis}, {Mueller}, {Olofsson}, {Patapis}, {Pye}, {Royer},
  {Serabyn}, {Scheithauer}, {Colina}, {van Dischoeck E.}, {Ostlin}, {T.}, \&
  {G}}]{malin2024}
{M{\^a}lin}, M., {Boccaletti}, A., {Baudoz}, P., {et~al.} 2024, \aap submitted

\bibitem[{{Mathis} {et~al.}(1977){Mathis}, {Rumpl}, \&
  {Nordsieck}}]{mathis1977}
{Mathis}, J.~S., {Rumpl}, W., \& {Nordsieck}, K.~H. 1977, \apj, 217, 425

\bibitem[{Matthews {et~al.}(2014)Matthews, Krivov, Wyatt, Bryden, \&
  Eiroa}]{matthews2014}
Matthews, B.~C., Krivov, A.~V., Wyatt, M.~C., Bryden, G., \& Eiroa, C. 2014,
  Protostars and Planets VI, 521

\bibitem[{{Mazoyer} {et~al.}(2014){Mazoyer}, {Boccaletti}, {Augereau},
  {Lagrange}, {Galicher}, \& {Baudoz}}]{Mazoyer2014}
{Mazoyer}, J., {Boccaletti}, A., {Augereau}, J.~C., {et~al.} 2014, \aap, 569,
  A29

\bibitem[{{Mittal} {et~al.}(2015){Mittal}, {Chen}, {Jang-Condell}, {Manoj},
  {Sargent}, {Watson}, \& {Lisse}}]{mittal2015}
{Mittal}, T., {Chen}, C.~H., {Jang-Condell}, H., {et~al.} 2015, \apj, 798, 87

\bibitem[{{Moore} {et~al.}(2023){Moore}, {Li}, {Hassenzahl}, {Nesvold}, {Naoz},
  \& {Adams}}]{moore2023}
{Moore}, N. W.~H., {Li}, G., {Hassenzahl}, L., {et~al.} 2023, \apj, 943, 6

\bibitem[{Müller {et~al.}(2018)Müller, Keppler, Henning, Samland, Chauvin,
  Beust, Maire, Molaverdikhani, Boekel, benisty, Boccaletti, Bonnefoy,
  Cantalloube, Charnay, Baudino, Gennaro, Long, Cheetham, Desidera, Feldt,
  Fusco, Girard, Gratton, Hagelberg, Janson, Lagrange, Langlois, Lazzoni, Ligi,
  Menard, Mesa, Meyer, Mollière, Mordasini, Moulin, Pavlov, Pawellek, Quanz,
  Ramos, Rouan, Sissa, Stadler, Vigan, Wahhaj, Weber, \& Zurlo}]{muller2018}
Müller, A., Keppler, M., Henning, T., {et~al.} 2018, Astronomy \&
  Astrophysics, 617, L2

\bibitem[{{Nesvold} {et~al.}(2017){Nesvold}, {Naoz}, \&
  {Fitzgerald}}]{nesvold2017}
{Nesvold}, E.~R., {Naoz}, S., \& {Fitzgerald}, M.~P. 2017, \apjl, 837, L6

\bibitem[{{Nguyen} {et~al.}(2021){Nguyen}, {De Rosa}, \& {Kalas}}]{nguyen2021}
{Nguyen}, M.~M., {De Rosa}, R.~J., \& {Kalas}, P. 2021, \aj, 161, 22

\bibitem[{{Olofsson} {et~al.}(2020){Olofsson}, {Milli}, {Bayo}, {Henning}, \&
  {Engler}}]{olofsson2020}
{Olofsson}, J., {Milli}, J., {Bayo}, A., {Henning}, T., \& {Engler}, N. 2020,
  \aap, 640, A12

\bibitem[{{Olofsson} {et~al.}(2016){Olofsson}, {Samland}, {Avenhaus},
  {Caceres}, {Henning}, {Mo{\'o}r}, {Milli}, {Canovas}, {Quanz}, {Schreiber},
  {Augereau}, {Bayo}, {Bazzon}, {Beuzit}, {Boccaletti}, {Buenzli}, {Casassus},
  {Chauvin}, {Dominik}, {Desidera}, {Feldt}, {Gratton}, {Janson}, {Lagrange},
  {Langlois}, {Lannier}, {Maire}, {Mesa}, {Pinte}, {Rouan}, {Salter},
  {Thalmann}, \& {Vigan}}]{Olofsson2016}
{Olofsson}, J., {Samland}, M., {Avenhaus}, H., {et~al.} 2016, \aap, 591, A108

\bibitem[{{Pan} {et~al.}(2016){Pan}, {Nesvold}, \& {Kuchner}}]{Pan2016}
{Pan}, M., {Nesvold}, E.~R., \& {Kuchner}, M.~J. 2016, \apj, 832, 81

\bibitem[{{Pearce} {et~al.}(2024){Pearce}, {Krivov}, {Sefilian}, {Jankovic},
  {L{\"o}hne}, {Morgner}, {Wyatt}, {Booth}, \& {Marino}}]{Pearce2024}
{Pearce}, T.~D., {Krivov}, A.~V., {Sefilian}, A.~A., {et~al.} 2024, \mnras,
  527, 3876

\bibitem[{{Pecaut} {et~al.}(2012){Pecaut}, {Mamajek}, \& {Bubar}}]{pecaut2012}
{Pecaut}, M.~J., {Mamajek}, E.~E., \& {Bubar}, E.~J. 2012, \apj, 746, 154

\bibitem[{{Perotti} {et~al.}(2023){Perotti}, {Christiaens}, {Henning},
  {Tabone}, {Waters}, {Kamp}, {Olofsson}, {Grant}, {Gasman}, {Bouwman},
  {Samland}, {Franceschi}, {van Dishoeck}, {Schwarz}, {G{\"u}del}, {Lagage},
  {Ray}, {Vandenbussche}, {Abergel}, {Absil}, {Arabhavi}, {Argyriou},
  {Barrado}, {Boccaletti}, {Caratti o Garatti}, {Geers}, {Glauser},
  {Justannont}, {Lahuis}, {Mueller}, {Nehm{\'e}}, {Pantin}, {Scheithauer},
  {Waelkens}, {Guadarrama}, {Jang}, {Kanwar}, {Morales-Calder{\'o}n},
  {Pawellek}, {Rodgers-Lee}, {Schreiber}, {Colina}, {Greve}, {{\"O}stlin}, \&
  {Wright}}]{perotti2023}
{Perotti}, G., {Christiaens}, V., {Henning}, T., {et~al.} 2023, \nat, 620, 516

\bibitem[{Phillips {et~al.}(2020)Phillips, Tremblin, Baraffe, Chabrier, Allard,
  Spiegelman, Goyal, Drummond, \& Hébrard}]{phillips2020}
Phillips, M.~W., Tremblin, P., Baraffe, I., {et~al.} 2020, Astronomy \&
  Astrophysics, 637, A38

\bibitem[{{Rodet} {et~al.}(2017){Rodet}, {Beust}, {Bonnefoy}, {Lagrange},
  {Galli}, {Ducourant}, \& {Teixeira}}]{rodet2017}
{Rodet}, L., {Beust}, H., {Bonnefoy}, M., {et~al.} 2017, \aap, 602, A12

\bibitem[{{Rouan} {et~al.}(2000){Rouan}, {Riaud}, {Boccaletti}, {Cl{\'e}net},
  \& {Labeyrie}}]{rouan2000}
{Rouan}, D., {Riaud}, P., {Boccaletti}, A., {Cl{\'e}net}, Y., \& {Labeyrie}, A.
  2000, \pasp, 112, 1479

\bibitem[{{Smith} \& {Terrile}(1984)}]{smith1984}
{Smith}, B.~A. \& {Terrile}, R.~J. 1984, Science, 226, 1421

\bibitem[{{Strubbe} \& {Chiang}(2006)}]{strubbe2006}
{Strubbe}, L.~E. \& {Chiang}, E.~I. 2006, \apj, 648, 652

\bibitem[{{Th{\'e}bault} \& {Augereau}(2007)}]{thebault2007}
{Th{\'e}bault}, P. \& {Augereau}, J.~C. 2007, \aap, 472, 169

\bibitem[{Thebault \& Kral(2019)}]{thebault2019}
Thebault, P. \& Kral, Q. 2019, Astronomy \& Astrophysics, 626, A24

\bibitem[{{Vandenbussche} {et~al.}(2004){Vandenbussche}, {Dominik}, {Min}, {van
  Boekel}, {Waters}, {Meeus}, \& {de Koter}}]{Vandenbussche2004}
{Vandenbussche}, B., {Dominik}, C., {Min}, M., {et~al.} 2004, \aap, 427, 519

\bibitem[{{Wright} {et~al.}(2023){Wright}, {Rieke}, {Glasse}, {Ressler},
  {Garc{\'\i}a Mar{\'\i}n}, {Aguilar}, {Alberts}, {{\'A}lvarez-M{\'a}rquez},
  {Argyriou}, {Banks}, {Baudoz}, {Boccaletti}, {Bouchet}, {Bouwman}, {Brandl},
  {Breda}, {Bright}, {Cale}, {Colina}, {Cossou}, {Coulais}, {Cracraft}, {De
  Meester}, {Dicken}, {Engesser}, {Etxaluze}, {Fox}, {Friedman}, {Fu},
  {Gasman}, {G{\'a}sp{\'a}r}, {Gastaud}, {Geers}, {Glauser}, {Gordon},
  {Greene}, {Greve}, {Grundy}, {G{\"u}del}, {Guillard}, {Haderlein},
  {Hashimoto}, {Henning}, {Hines}, {Holler}, {Detre}, {Jahromi}, {James},
  {Jones}, {Justtanont}, {Kavanagh}, {Kendrew}, {Klaassen}, {Krause},
  {Labiano}, {Lagage}, {Lambros}, {Larson}, {Law}, {Lee}, {Libralato}, {Lorenzo
  Alverez}, {Meixner}, {Morrison}, {Mueller}, {Murray}, {Mycroft}, {Myers},
  {Nayak}, {Naylor}, {Nickson}, {Noriega-Crespo}, {{\"O}stlin}, {O'Sullivan},
  {Ottens}, {Patapis}, {Penanen}, {Pietraszkiewicz}, {Ray}, {Regan},
  {Roteliuk}, {Royer}, {Samara-Ratna}, {Samuelson}, {Sargent}, {Scheithauer},
  {Schneider}, {Schreiber}, {Shaughnessy}, {Sheehan}, {Shivaei}, {Sloan},
  {Tamas}, {Teague}, {Temim}, {Tikkanen}, {Tustain}, {van Dishoeck},
  {Vandenbussche}, {Weilert}, {Whitehouse}, \& {Wolff}}]{wright2023}
{Wright}, G.~S., {Rieke}, G.~H., {Glasse}, A., {et~al.} 2023, \pasp, 135,
  048003

\bibitem[{{Wyatt}(2008)}]{wyatt2008}
{Wyatt}, M.~C. 2008, \araa, 46, 339

\end{thebibliography}

\begin{appendix}

\section{Computation of $Q_{abs}$ at $\lambda$ = 1.27\,mm for grains larger than 10\,\mic}
\begin{figure}[!htb]
	   	\centering
	   	\includegraphics[width=9cm]{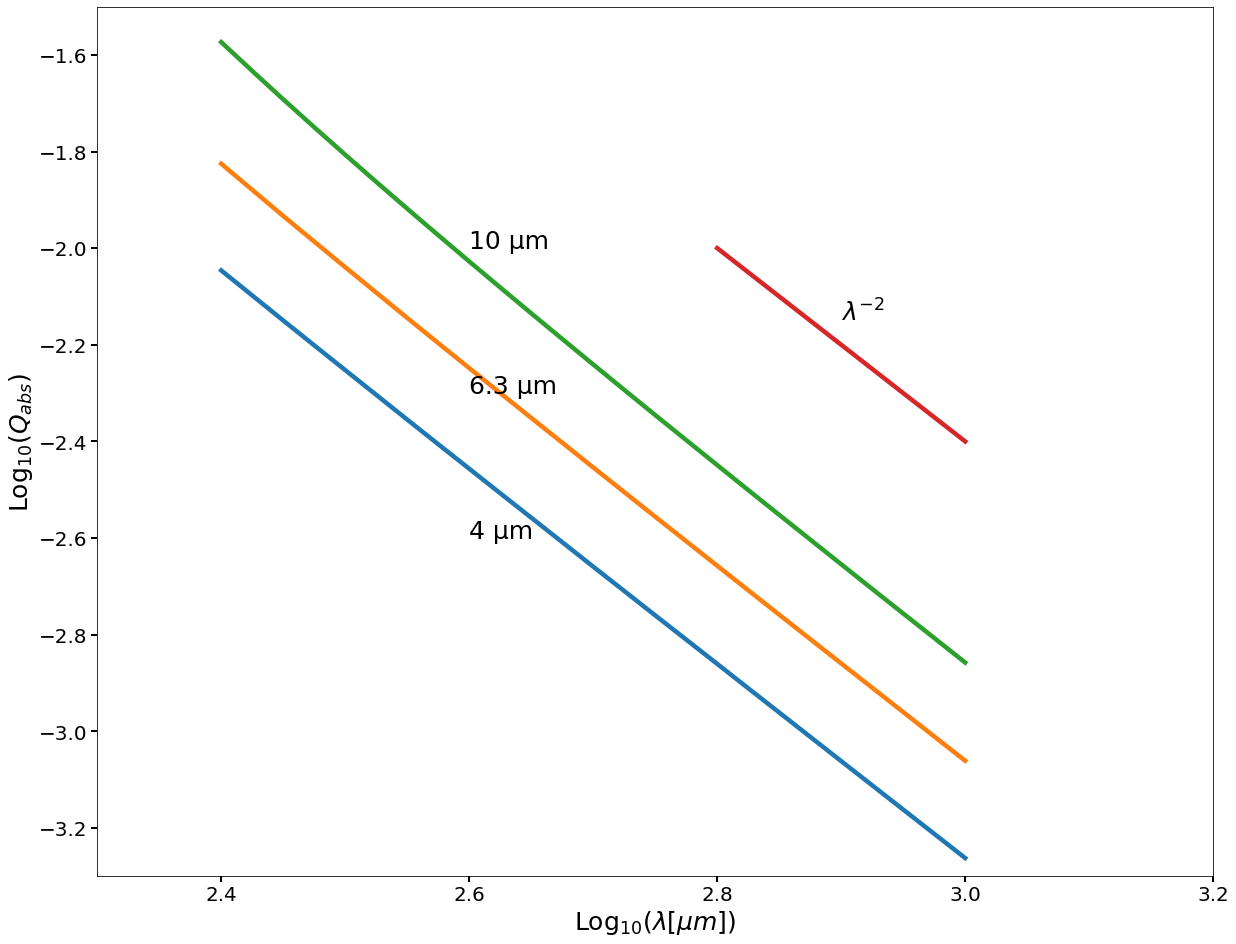}
	   	\caption{Behavior of $Q_{abs}$  vs $\lambda$ in the range $\lambda$ = 0.25 to 1 mm for
	   	three sizes of silicate grains : 4, 6.3 and 10 µm. }
	   	\label{fig:A1}
   	\end{figure}

The optical properties of silicate that we used \citep{laor1993} are limited 
in size to a$_{max}$ = 10 \mic and in wavelengths to $\lambda_{max}$ = 1 mm.
 To extrapolate  Q$_{abs}$ at the wavelength $\lambda_0 =$ 1.27 mm, we observe that 
 for large Silicate grains the behavior of  $Q_{abs}$  vs $\lambda$ follows remarkably well a
 power law with exponent -2, as shown on Fig. \ref{fig:A1} for a few values of $a$ and in the 
 the range of wavelengths 250 - 1000\,\mic. 

\begin{figure}[!htb]
	   	\centering
	   	\includegraphics[width=9cm]{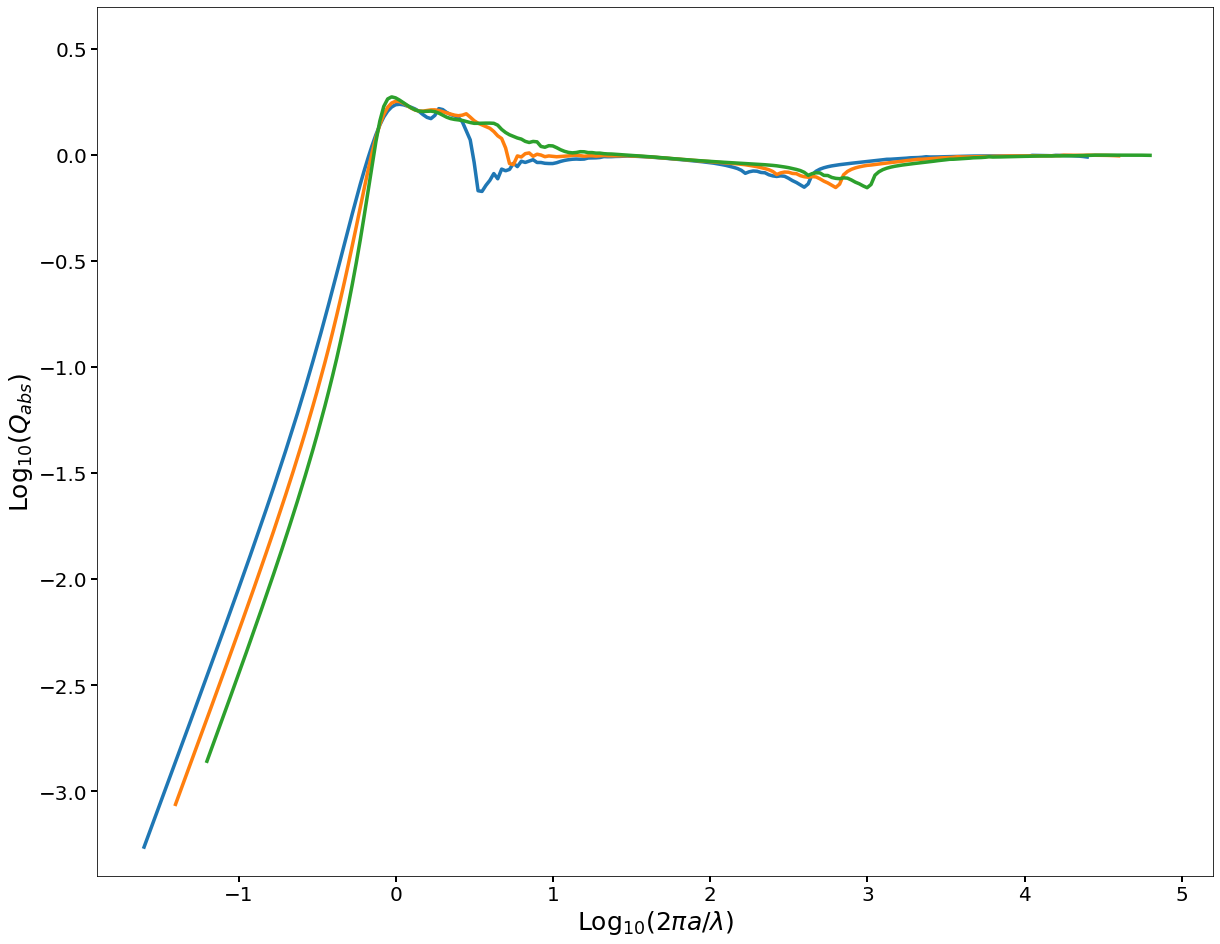}
	   	\caption{Behavior of $Q_{abs}$  vs $2 \pi a / \lambda$ in the range $\lambda$ = 0.25 to 1 mm, 
     for three sizes of silicate grains : 4, 6.3 and 10 µm, using the same color code as for Fig. \ref{fig:A1}. }
	   	\label{fig:A2}
   	\end{figure}

As regards the extrapolation of  $Q_{abs}(1270)$ when grain size is larger than 10\,\mic, 
we show in Fig. \ref{fig:A2} that the hypothesis that  $Q_{abs}(a, \lambda)$  depends only 
on the quantity $2 \pi a / \lambda$ is already almost verified for Silicate grains of 4, 6.3 and 10\,\mic,
so that using this result for any size larger than 10\,\mic is legitimate. Using then 
$Q_{abs}(10\,\mic, \lambda)$ as a template, the equation to compute $Q_{abs}$ for any grain 
larger than 10 \mic becomes : \\
 $Q_{abs}(a, \lambda)$ =  
 $Q_{abs}( 10,  a \times \lambda_0 / 10 )$.

\section{Comparison with models in worst cases}  
\begin{figure*}[!htb]
	   	\centering
	   	\includegraphics[width=18cm]{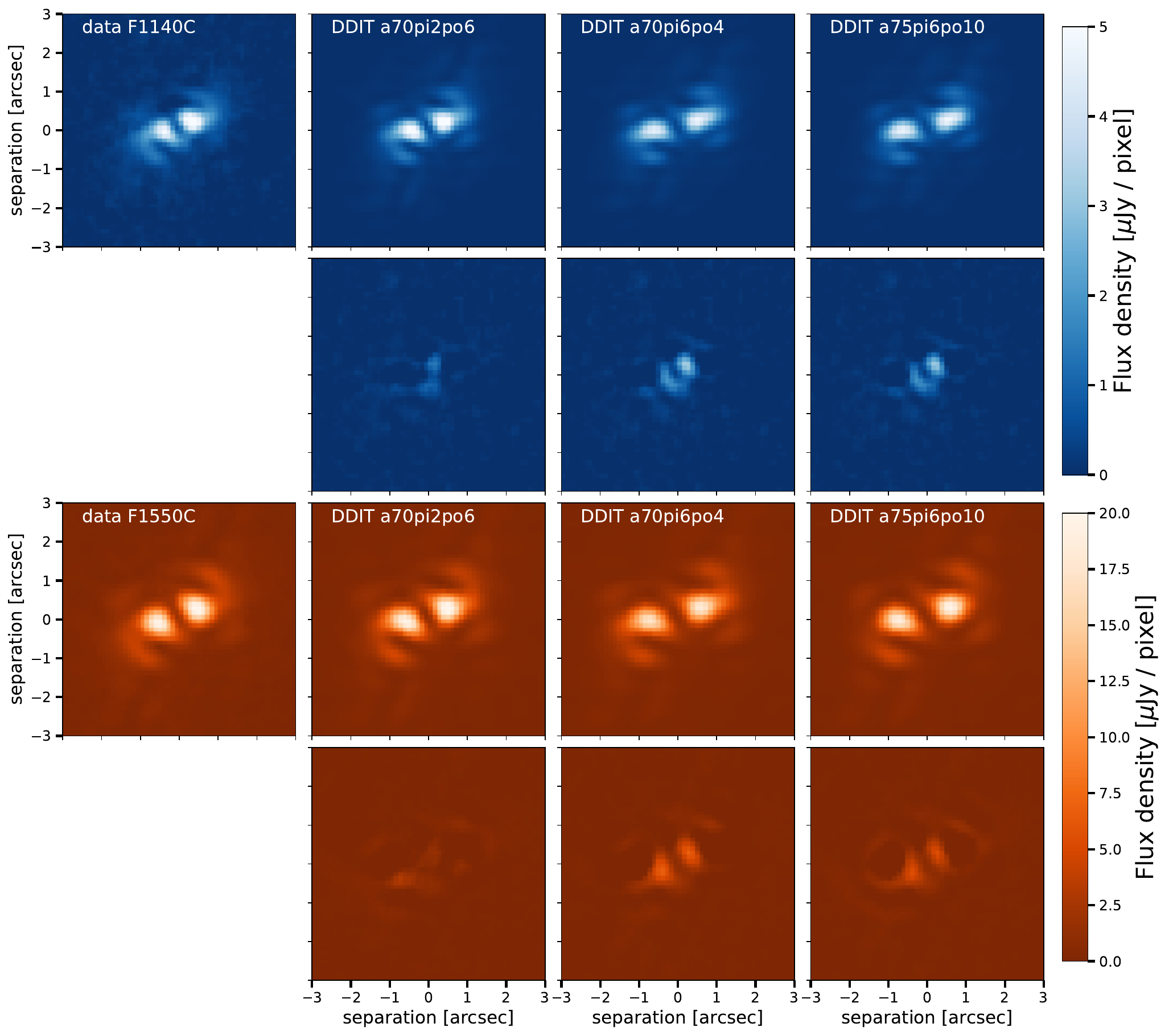}
	   	\caption{Comparison between observed images and DDiT+ models in cases where residues are large. 
     The residues are below the simulated images. 
     The first column is the observation, the second the best case, and the two last ones show, two examples of bad cases.}
	   	\label{fig:worstfits}
   	\end{figure*}

Fig. \ref{fig:worstfits} displays two cases of parameters for which the DDiT+ models provide clearly a 
bad fit to the observations, as compared to the best model.

\section{Appearance of the whole disk in the best case}
\begin{figure*}[!htb]
	   	\centering
	   	\includegraphics[width=16cm]{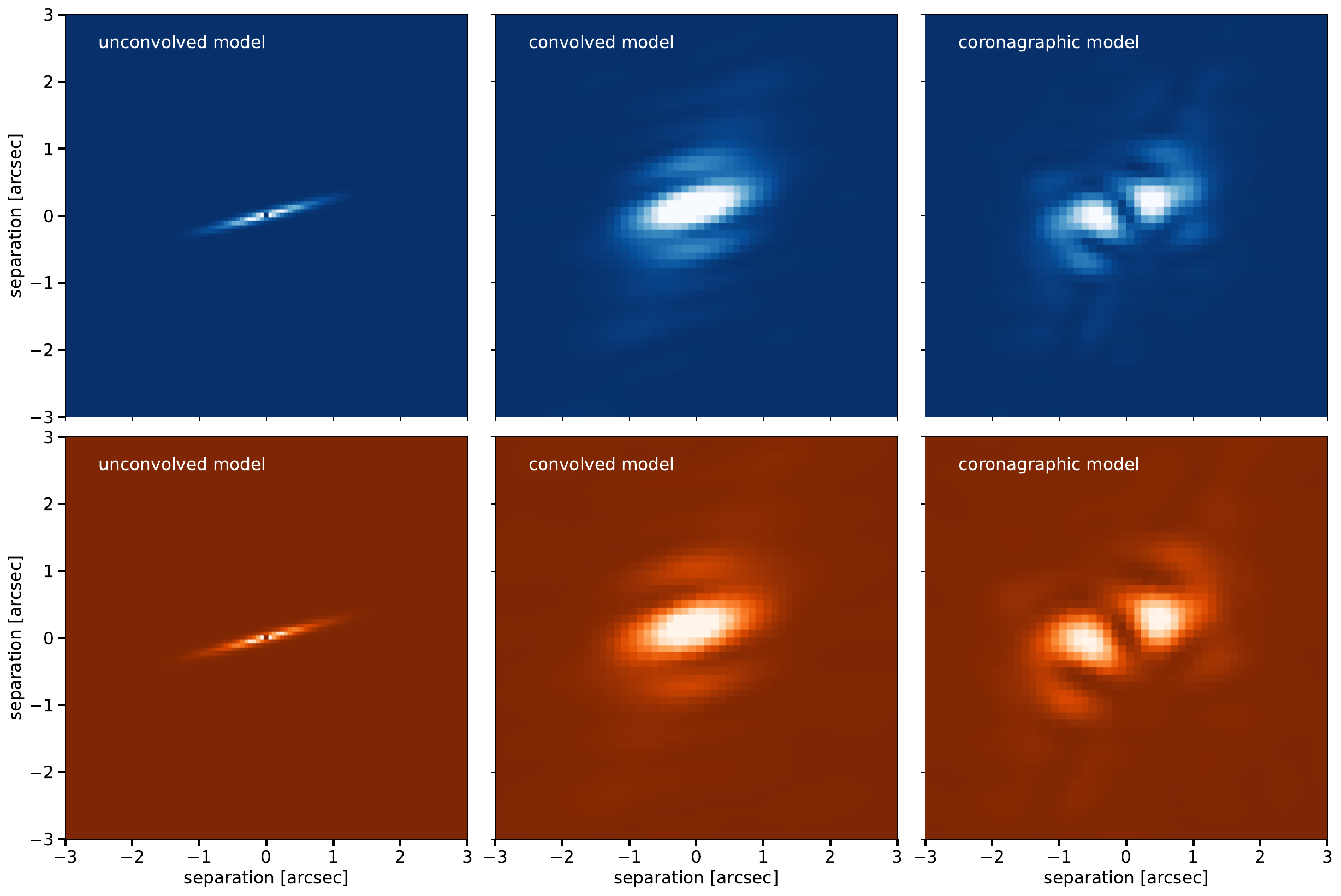}
\caption{Illustration of the different steps of the simulation of synthetic disk images for the two filters (F1140C at the top, F1550C at the bottom). From left to right: DDiT+ disk model, non-coronagraphic image, coronagraphic image). 
} 
	   	\label{fig:coronomodel}
   	\end{figure*}

Fig. \ref{fig:coronomodel} displays the appearance of the best DDiT+ models at the various stages of the numerical simulations.

\section{Spitzer spectrum: indication of forsterite grains ?}
In the IRS-Spitzer spectrum of HD106906 extracted from the CASSIS Atlas (https://cassis.sirtf.com/atlas/), we identified (circle) two emission features at  23.7 and 34.0 \mic that likely are  characteristic of forsterite crystalline silicate \citep{Vandenbussche2004}. 

	\begin{figure*}[!htb]
	   	\centering

	   	\includegraphics[width=9cm]{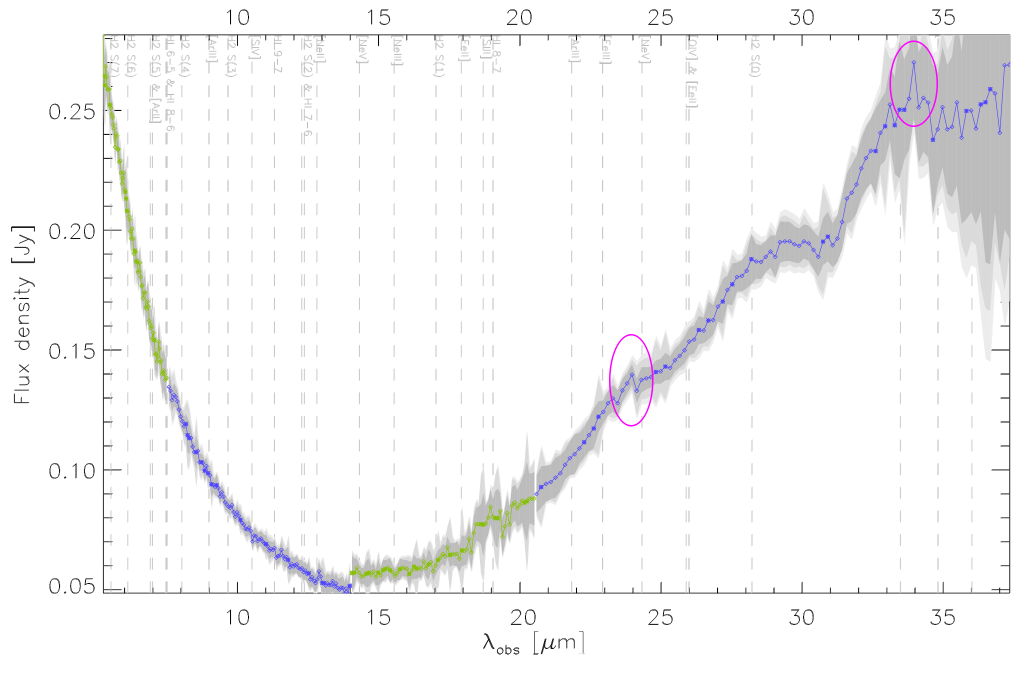}
\caption{IRS-Spitzer spectrum of HD106906 extracted from the CASSIS Atlas (https://cassis.sirtf.com/atlas/). We marked (ovals) two emission features at  23.7 and 34.0 \mic that are likely characteristic of forsterite (crystalline silicate).  
} 
	   	\label{fig:A5}
   	\end{figure*}

\end{appendix}
\end{document}